\documentclass{andp2012}
\usepackage[english]{babel}
\usepackage{url}
\usepackage{graphicx}
\title{Information theory for fields}
\author[T. A. En{\ss}lin]{Torsten A. En{\ss}lin\footnote{Corresponding author\quad E-mail:~\textsf{ensslin@mpa-garching.mpg.de}\\ Max Planck Institute for Astrophysics, Karl-Schwarzschild-Str. 1, 85741 Garching, Germany}}

\begin{abstract}
A physical field has an infinite number of degrees of freedom since it has a field value at each location of a continuous space. Therefore, it is impossible to know a field from finite measurements alone and prior information on the field is essential for field inference. An information theory for fields is needed to join the measurement and prior information into probabilistic statements on field configurations. Such an information field theory (IFT) is built upon the language of mathematical physics, in particular on field theory and statistical mechanics. IFT permits the mathematical derivation of optimal imaging algorithms, data analysis methods, and even computer simulation schemes. The application of IFT algorithms to astronomical datasets provides high fidelity images of the Universe and facilitates the search for subtle statistical signals from the Big Bang. The concepts of IFT might even pave the road to novel computer simulations that are aware of their own uncertainties.
\end{abstract}
\shortabstract
\begin{document}
\maketitle
\section{Information field theory}\label{sec:IFT}

\begin{figure*}
\centering
\includegraphics[width=0.7\linewidth]{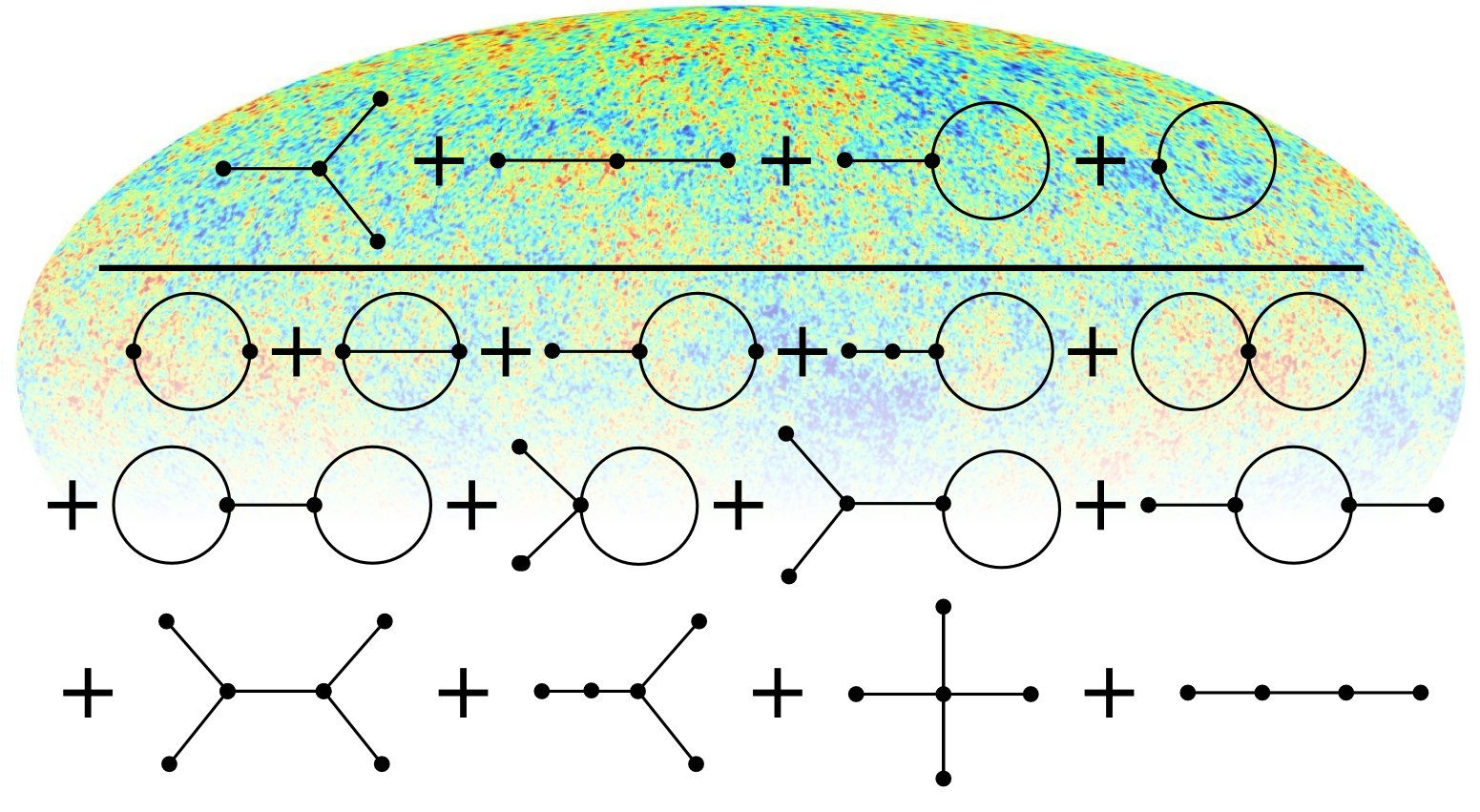}
\caption{Usage of Feynman diagrams in IFT. An estimator for the level of non-Gaussianity of a field is displayed on top of temperature fluctuations of the CMB, to which this estimator can be applied to study the Early Universe. Dots at the end of lines represent the information source $j$ provided by the data, lines the information propagator $D$, and internal vertices with three or four lines cubic and quartic terms of the information Hamiltonian in Eq.~\ref{eq:fnl}, respectively.}
\label{fig:fnl}
\end{figure*}

\subsection{Aim}\label{sec:aim}

\textit{Information field theory} (IFT) \cite{2009PhRvD..80j5005E, 2013AIPC.1553..184E, 2014AIPC.1636...49E, wiki:xxx} is information theory for fields. It uses methods from quantum and statistical field theories for the inference of field properties from data. IFT permits to design optimal algorithm to recover unknown or partially known signal fields from noisy, incomplete, and otherwise corrupted measurements given that certain statistical knowledge on the physical nature of the field is available. 

To this end, IFT uses Bayesian probability theory for fields. Bayesian probabilities quantify how much one should believe in a statement being true, a condition being met, or an event to occur. They are bookkeeping devices for evidences. They are not necessarily expressing physical frequencies of events. In IFT, logarithmic probabilities will turn out to be the currency to convert and combine different forms of information, both from measurements and also from prior knowledge and assumptions. 

This article introduces into IFT with the aim to put the reader into a position to follow the more specialized literature on IFT, to understand and use existing IFT methods, or to develop those adapted to his or her own needs.  Basic knowledge of the reader on field theory and thermodynamics is assumed, although, most concepts used are briefly introduced. 

\subsection{Structure of the work}

The structure of the work is the following. In this Sect.~\ref{sec:IFT}, the problem of field inference is approached with Bayesian probability theory. In Sect.~\ref{sec:free}, the free theory of IFT is developed using an illustrative example of a diffusion process driven by white noise and observed at some point in time.
In Sect.~\ref{sec:interactingIFT}, this is then extended into the non-linear regime, leading to interacting IFT in wich the field estimate is not a linear function of the data any more. Sect.~\ref{sec:applications} discusses a number of existing and envisaged IFT algorithms and their numerical implementation. An outlook to future developments in IFT are given in the final Sect.~\ref{sec:outlook}

\subsection{Physical fields}

Physical fields like the electromagnetic field, the gravitational field, or even macroscopic ones like the atmospheric temperature generally exhibit different values at different locations in space and time. Knowing the configuration of a field, \textit{i.e.} knowing the field values at all locations, can be of tremendous scientific, technological or economical value. Unfortunately, this is impossible in general, as there are more locations to be known (often infinitely many) than available measurement data that provide mathematical constraints on the possible field configurations. Thus, inferring a field is an under-constrained problem in general. For a given dataset obtained by an instrument probing the field, there can be an infinite number of potential field configurations that are fully consistent with the data. How to choose among them?

Many of those configurations will be implausible, \textit{e.g.} imagine a radio telescope observing the cosmic microwave background (CMB). The telescope is a measurement device that integrates the field over some sub-volume of space and provides this integral. Then the integrated field in this volume is known, but how this amount of integrated field is distributed among the different locations is not determined. Among all configurations that obey this measurement constraint, very rough ones (i.e.\ ones that exhibit completely different values at the different locations) are possible and outnumber the also possible smooth ones largely. There are just many more ways for a field to be jagged than ways to be smooth. Nevertheless, for most physical fields, the smooth field configurations would be regarded as far more plausible. Rapid spatial changes of a field are less likely as they either cost energy (think of an electrical potential field) or are erased rapidly by the field dynamics (think of temperature jumps in a turbulent or diffusive atmosphere). Thus, among all configurations consistent with the data, smooth ones should be favored. But how smooth should they be? And how should this extra information on smoothness be inserted into the field inference?

\subsection{Smoothness and correlations}

The first question requires a specification of the concept of smoothness. A field can be regarded to be smooth, if field values at nearby locations are similar. Thus, knowing the field at one location implies some constraints about the possible values at nearby positions. Such knowledge is best represented in terms of correlations between the different locations. Smoothness of a field $ \varphi $ is therefore well characterized by the two point correlation
\begin{equation}\label{eq:Phi}
\Phi(x,y) = \langle \varphi(x)\, \varphi(y)\rangle_{(\varphi)}
\end{equation}
of locations $ x $ and $ y $, assuming for simplicity here that the field is fluctuating around zero such that its expectation value vanishes, $\langle \varphi \rangle_{(\varphi)}=0$. 

We denote with $\langle f(a,b) \rangle_{(a|b)}= \int \mathrm{d}a\, f(a,b) \, \mathcal{P}(a|b)$ the expectation of $f(a,b)$ averaged over the conditional probability density $ \mathcal{P}(a|b)$ for the variable $a$ given $b$. We postpone the question how such integrals are calculated for $a$ being a field. We denote probabilities with $P\in [0,1]$, and probability densities with $\mathcal{P}\in [0,\infty)$. In the following, we often drop the word density for briefness. Thus, the expectation value in Eq.~\ref{eq:Phi} should be performed over all plausible field configurations $\varphi$ weighted with their appropriate probability densities $\mathcal{P}(\varphi)$. Such probabilities will be constructed later on.

For fields that have a spatially homogeneous statistics, this correlation function is only a function of the difference of the locations, and additionally for statistically isotropic fields, it depends only on the absolute distance $r= |x-y|$, $\Phi(x,y) = C_\varphi(|x-y|) $, with $C_\varphi(r)$ the radial two-point correlation function of the field $\varphi$. For example, the density field of matter in the cosmos on large scales is assumed to be the result of a statistically isotropic and homogeneous random process. For this reason, the CMB shows temperature fluctuations with isotropic correlation statistics (see Fig.~\ref{fig:fnl}).

The correlation function  $C_\varphi(r)$ reflects the roughness of a statistical homogeneous and isotropic field. It is maximal at zero lag $r=x-y$ and falls towards zero typically after a correlation length
\begin{equation}
\lambda = \int_0^\infty  dr \, \frac{C_\varphi(r)}{C_\varphi(0)}, 
\end{equation}
meaning that field values at locations separated by $r\gg \lambda$ are largely uncorrelated. On scales smaller than $\lambda$ field values are strongly correlated and the field appears smooth on such distances. Knowledge of the value of a correlated field at some location therefore makes a statement about the field values at nearby locations, within the correlation length. They should be similar, in particular, if they are closer by.

The knowledge and usage of such correlations is essential for IFT, as it provides an infinite number of statements on the field, connecting all pairs of points in a statistical manner. This helps to tame the infinite number of degrees of freedom of an unknown field. Traditional smoothness regularization, like the suppression of strong gradients or curvature of the field, are naturally included in this language, as we will see in Sect.~\ref{sec:regularization}.

A great effort in an IFT analysis of a signal inference problem should therefore go into determining the correlation structure of the field. Ideally, it is derived from a physical line of reasoning, as being presented for the example developed later on in in Sect.~\ref{sec:simplescenario}.
In case such prior information on the field correlation is not available, it might be extracted from the data itself, as discussed in Sect.~\ref{sec:critical}.

Assuming for now that we have some usable knowledge on the field smoothness the second question remains: How to incorporate this knowledge into the inference?

\begin{figure*}
\centering
\includegraphics[width=0.7\linewidth]{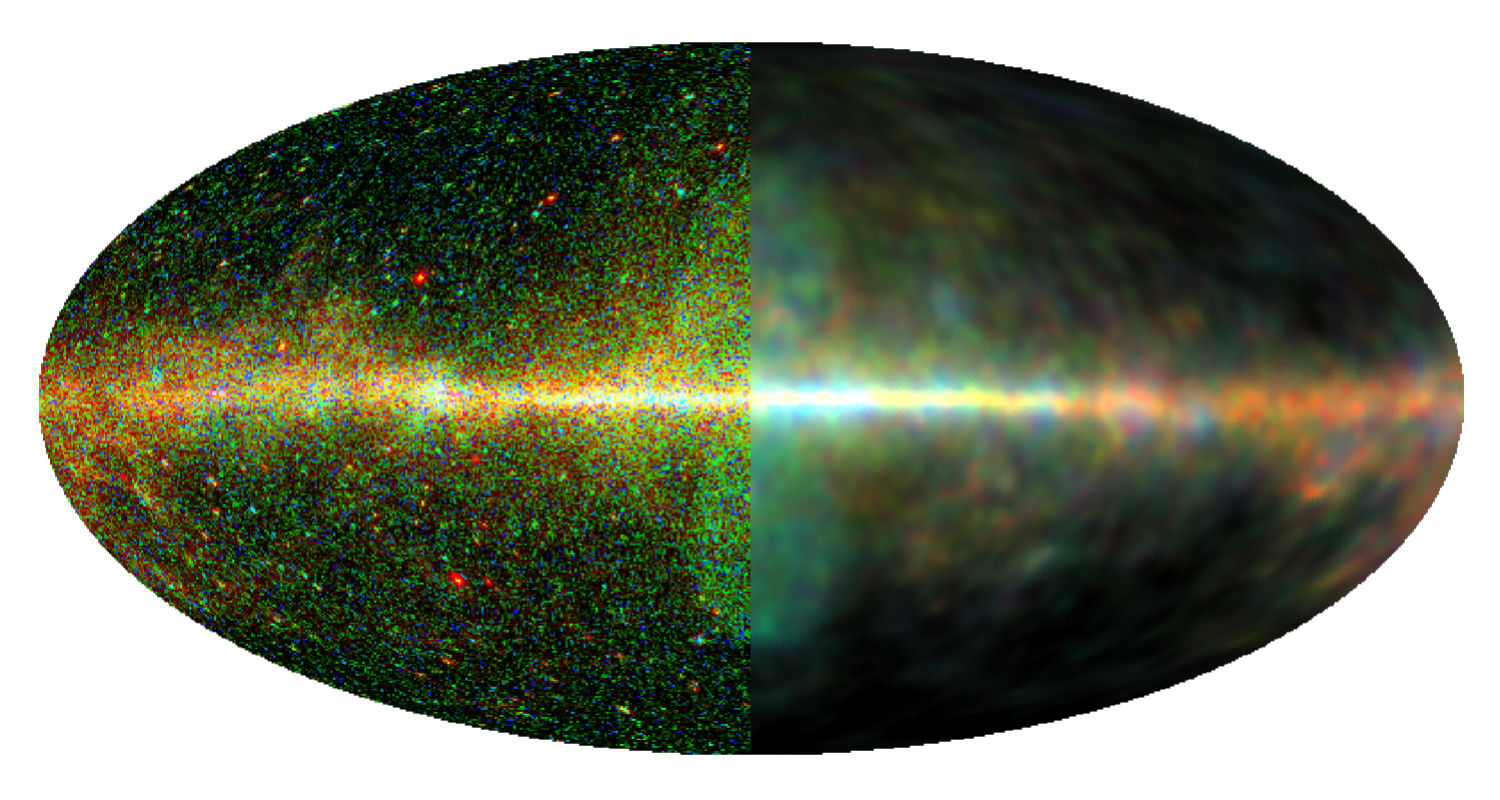}
\caption{The gamma-ray sky as seen by the Fermi satellite. Left half: original photon counts, intensity is encoded in the brightness, photon energy in the color (red $\sim$ 1 GeV, blue $\sim$ 100 GeV). Right half: diffuse emission component reconstructed by the D$^3$PO algorithm, which takes out the shot noise and the point sources.}
\label{fig:fermi-diffuse}
\end{figure*}

\subsection{Bayes and statistical physics}

Bayes' theorem conveniently answers this question. Assume we are interested in a field $\varphi: \Omega \rightarrow\mathbb{R}$, which is defined on some $u$-dimensional space $\Omega \subseteq \mathbb{R}^u$ and we have obtained some measurement data
\begin{equation}\label{eq:d}
d = R(\varphi) + n
\end{equation}
on it, where $R(\varphi) = \langle d \rangle_{(d|\varphi)}$ is the deterministic response of the instrument to the field, and $n = d- R(\varphi)$ denotes any stochastic contribution to the outcome, called the measurement noise. 
For example, a linearly integrating instrument is described by
\begin{equation}\label{eq:d_comp}
R_i(\varphi) = \int_{\Omega}\!\! \mathrm{d}x \, r_i(x)\, \varphi(x) \equiv (R\,\varphi)_i,
\end{equation}
where $i$ indexes the different measurements and $r_i(x)$ how those respond to the field values at different locations $x$.

The measurement equation, Eq.~\ref{eq:d}, in combination with the noise statistics $\mathcal{P}(n|\varphi)$ for a given field configuration $\varphi$, defines the \textit{likelihood} to obtain the data $d$,
\begin{eqnarray}
\mathcal{P}(d|\varphi) &=& \int \mathcal{D}n\, \delta(d-R(\varphi) -n)\, \mathcal{P}(n|\varphi)\nonumber\\
& =& \mathcal{P}(n= d- R(\varphi)|\varphi).
\end{eqnarray}
This likelihood, in combination with the data, contains the full information the measurement is providing. A likelihood function, which is as accurate as possible, is essential to extract the full information content of a data set. In case some properties of the measurement process are not precisely known, their determination has to be made part of the inference problem.  This can turn an otherwise simple linear signal inference task into a complex instrumental self-calibration problem. For the moment, such complications are assumed to be absent.

Since the data is only providing finite dimensional constraints on the many possible field configurations, whereas the signal field has infinite number of degrees of freedom, it is not sufficient to fully specify the field. The information from the data has to be combined with prior knowledge, for example on the field correlations as encoded in $C_\varphi(r)$, to provide a \textit{posterior  probability} density on the field $\mathcal{P}(\varphi|d)$ (given the data and other information), from which field estimates could be obtained. One such estimate would be the posterior mean field
\begin{equation}
	m = \langle \varphi \rangle_{(\varphi|d)} = 
	\int \mathcal{D}\varphi \, \mathcal{P}(\varphi|d)\, \varphi .
\end{equation}
Here, an integration over all field configurations is necessary, a technicality we postpone until Sect.~\ref{sec:infinity}.

Bayes' theorem states that this posterior $\mathcal{P}(\varphi|d) $, which summarizes our knowledge on the field after the data is taken, is proportional to the product of the data likelihood $ \mathcal{P}(d|\varphi) $ for a given field configuration and its \textit{prior probability} $ \mathcal{P}(\varphi) $, which summarizes our a-priori knowledge on the field (\textit{e.g.} smooth configurations are more plausible than rough ones). Bayes' theorem reads
\begin{equation}
\mathcal{P}(\varphi|d) = \frac{\mathcal{P}(d|\varphi) \,\mathcal{P}(\varphi) }{\mathcal{P}(d) },\label{eq:Bayes_theorem}
\end{equation}
where the normalization of the posterior is provided by the so-called \textit{evidence}
\begin{equation}
\mathcal{P}(d) = \int \mathcal{D} \varphi \,\mathcal{P}(d|\varphi) \,\mathcal{P}(\varphi).
\end{equation}

In other words, the posterior is the joint probability of data and field $ \mathcal{P}(d,\varphi) = \mathcal{P}(d|\varphi) \,\mathcal{P}(\varphi)$ evaluated for the observed data $d=d_\mathrm{obs}$, and normalized such that it becomes a proper probability density gain, $ \int \mathcal{D} \varphi \,\mathcal{P}(\varphi|d=d_\mathrm{obs})=1$. 

Formally, Bayes theorem can be brought into a form resembling the Boltzmann distribution of statistical mechanics,\footnote{The temperature usually appearing in thermodynamics has been set to $T=1$  here.}
\begin{equation}\label{eq:BayesH}
\mathcal{P}(\varphi|d) = \frac{\exp(-\mathcal{H}(d,\varphi)) }{\mathcal{Z}(d) },
\end{equation}
where we have introduced the information Hamiltonian
\begin{equation}
\mathcal{H}(d,\varphi) = -\ln \mathcal{P}(d,\varphi),
\end{equation}
the negative logarithm of the joint probability of data and field.\footnote{The dimension of the probability density are ignored in this definition, as they have no influence on any result in IFT. }
The \textit{partition function}
\begin{equation}
\mathcal{Z}(d) = \int \mathcal{D} \varphi \, \exp(-\mathcal{H}(d,\varphi))
\end{equation}
is just another name for the evidence $\mathcal{P}(d)$. 

The statistical mechanics version of Bayes theorem, Eq.~\ref{eq:BayesH}, is a mere rewriting. The introduced information Hamiltonians (aka negative log-probabilities, surprises, or simply information) has the nice property that information appears as an additive quantity, since
\begin{equation}
	\mathcal{H}(d,\varphi) = \mathcal{H}(d|\varphi)  + \mathcal{H}(\varphi) 
\end{equation}
is a direct consequence of the product rule of probabilities, $ \mathcal{P}(d,\varphi) = \mathcal{P}(d|\varphi) \,\mathcal{P}(\varphi)$.
In particular the fusion of the information provided by two independent measurements, with the joint data vector $d=(d_1,d_2)$, is simply performed by addition of their information Hamiltonians,
\begin{equation}
\mathcal{H}(d|\varphi) = \mathcal{H}(d_1|\varphi)  + \mathcal{H}(d_2|\varphi) .
\end{equation}

The rewriting of Bayes' theorem for fields in the language of statical mechanics does not solve the problem to extract useful statements form the field posterior $\mathcal{P}(\varphi|d)$. However, it makes use of the many methods, which have been developed to treat statistical and quantum field theories. Consequently, we call the treatment of field inference problems in this language  \textit{information field theory}.\footnote{The term \textit{Bayesian field theory} was proposed  originally by Lemm \cite{Lemm2003} for field inference. This term, however, does not follow the tradition to name field theories after subjects and not people. We do not talk about Maxwell, Einstein, or Feynman field theory, but about electromagnetic, gravitational, and quantum field theories.}

\subsection{Expectation values}

One of the strengths of the field theoretical language is that it provides tools to calculate expectation values. For example, the partition function can be turned into a moment generating functional by extending it to
\begin{equation}
\mathcal{Z}(d,j) = \int \mathcal{D}\varphi \,  \exp(-\mathcal{H}(d,\varphi)+j^\dagger \varphi),
\end{equation}
where $j$ is a moment generating source term and
\begin{equation}
j^\dagger \varphi =\int_\Omega \!\!\mathrm{d}x\, j^*(x) \, \varphi(x)
\end{equation}
is the usual scalar product for functions. Any posterior expectation value of moments of field values can be obtained from the partition function via
\begin{equation}
\langle \varphi(x_1)\ldots \varphi(x_n)\rangle_{(\varphi|d)}=
\left.
\frac{1}{\mathcal{Z}(d,j)}\,
\frac{\partial^n \mathcal{Z}(d,j) }{\partial j^*(x_1) \ldots \partial j^*(x_n)}
\right|_{j=0}.
\end{equation}
The generation of the posterior mean field 
\begin{eqnarray}
m(x) &=& \langle \varphi(x) \rangle_{(\varphi|d)} \nonumber\\
&=& 
\left.
\frac{1}{\mathcal{Z}(d,j)}\,
\frac{\partial \mathcal{Z}(d,j) }{\partial j^*(x)}
\right|_{j=0}
=
\left.
\frac{\partial \ln \mathcal{Z}(d,j) }{\partial j^*(x)}
\right|_{j=0}
\end{eqnarray}
suggests the introduction of the so-called cummulant or connected moments via
\begin{equation}
\langle \varphi(x_1)\ldots \varphi(x_n)\rangle^{(\mathrm{c})}_{(\varphi|d)}=
\left.
\frac{\partial^n \ln \mathcal{Z}(d,j) }{\partial j^*(x_1) \ldots \partial j^*(x_n)}
\right|_{j=0}.
\end{equation}
The logarithmic partition function $ \ln \mathcal{Z}(d,j)$, which appears hereby, is known to be given by the sum of all singly connected Feynman diagrams without free vertices (see Fig.~\ref{fig:fnl} for such diagrams). This is the reason for the term \textit{connected moments}. The connected and unconnected moments are related, for example the posterior field dispersion
\begin{eqnarray}
D(x,y) &=& \langle \varphi(x) \,  \varphi(y)  \rangle^{(\mathrm{c})}_{(\varphi|d)}=
\left.
\frac{\partial^2 \ln \mathcal{Z}(d,j) }{\partial j^*(x) \partial j^*(y)}
\right|_{j=0}\nonumber \\
&=&
\left.
\frac{\partial }{\partial j^*(x)} \, \left(\frac{1}{ \mathcal{Z}(d,j)}\, \frac{\partial \mathcal{Z}(d,j) }{\partial j^*(y)} \right)
\right|_{j=0}\nonumber \\
&=&
\left.
\frac{1}{\mathcal{Z}}\frac{\partial^2 \mathcal{Z} }{\partial j^*(x) \partial j^*(y)}
-
\frac{1}{\mathcal{Z}^2}\frac{\partial \mathcal{Z} }{\partial j^*(x)}\,
\frac{\partial \mathcal{Z}}{\partial j^*(y)}
\right|_{j=0}\nonumber \\
&=&
\langle \varphi(x) \,\varphi(y)\rangle_{(\varphi|d)} -
  \,\varphi(y)\rangle_{(\varphi|d)}\,
  \varphi(y)\rangle_{(\varphi|d)}\nonumber
\\
&=& \langle [\varphi(x)-m(x)]\,[\varphi(y)-m(y)] \rangle_{(\varphi|d)},
\end{eqnarray}
is the second field moment corrected for the contribution of the first moment to it.

Of particular interest in practical applications are the first and second connected posterior moments of the field
that represent the posterior mean field $ m $ and the uncertainty covariance $ D $ around this mean. The mean is the best guess for the field under the $\mathcal{L}^2$-error loss function $\mathcal{L}^2=\langle ||\varphi-m||^2 \rangle_{(\varphi|d)}\equiv \langle (\varphi-m)^\dagger  (\varphi-m) \rangle_{(\varphi|d)} $ as a short calculation verifies:
\begin{eqnarray}
\frac{\partial \mathcal{L}^2}{\partial m} = 2\, \langle m-\varphi \rangle_{(\varphi|d)} = 0
%
\Rightarrow
m = \langle \varphi \rangle_{(\varphi|d)}.  \label{eq:L2mean}
\end{eqnarray}

\subsection{Maximum entropy}

Let us assume for a moment that the mean $ m = \langle \varphi \rangle_{(\varphi|d)} $ and the covariance  $ D =\langle \varphi \,  \varphi^\dagger  \rangle^{(\mathrm{c})}_{(\varphi|d)} \equiv  \langle (\varphi-m)\, (\varphi-m)^\dagger \rangle_{(\varphi|d)} $
is all what is known about a field, i.e. the data are this mean and covariance, $ d'=(m,\, D) $. Which probability distribution should then be used to represent the knowledge state? 

According to the maximum entropy principle \cite{1982ieee...70..939J}, the knowledge provided by this data is best summarized by a Gaussian probability distribution with this mean and covariance:
\begin{eqnarray}\label{eq:Gauss}
\mathcal{P}(\varphi|d')&=&\mathcal{G}(\varphi-m,\,D)\nonumber\\
&\equiv&
\frac{1}{\sqrt{|2\pi\,D|}}\,\exp\left[ -\frac{1}{2} (\varphi -m)^\dagger D^{-1} (\varphi -m) \right]
\end{eqnarray}

This is one of the lines of reasoning that is the basis of the frequent appearance of \textit{Gaussian processes}\footnote{A Gaussian process generates field configurations in such a way that the joint probability function of any set of field values evaluated at a set of distinct location is a multivariate Gaussian.} in IFT. Gaussian distributions can appear as minimal informative descriptions of field knowledge, only the first and second moments are needed to be known to specify those. 

The usage of a Gaussian field prior therefore does not necessarily imply that the field is following Gaussian statistics. Just the knowledge on the field statistics might be so limited that more elaborate priors can not be justified by the requirement to encode only the least amount of information consistent with all constraints into probabilities, the principle of maximum entropy. We recall, in Bayesian probability theory probabilities express knowledge on a quantity, not necessary the real, physical frequency of that quantity. If, however, the latter is known, it is also the correct probability to be used. We will encounter such a case  in Sec.~\ref{sec:simplescenario}.

\subsection{Infinities}\label{sec:infinity}

So far we have operated with fields, linear operators on fields like $R$ or $\Phi^{-1}$, and integrals over field configurations as if they were finite dimensional vectors, matrices, and integrals. One can imagine to perform such operations with pixelized versions of fields, operators, and the like, with the pixelization taken so fine that further refinements would not make any difference to calculated expectation values anymore. In order to approach the realm of fields, the limit of an infinite number of pixels has to be taken. Clearly, some quantities used above become infinite or vanish in this limit, like the determinant of a covariance $ D $ in Eq.~\ref{eq:Gauss}. However, these exploding terms are usually auxiliary expressions, ensuring the normalization of a probability distribution (with respect to a given pixelization) and are not of fundamental interest. What needs to stay finite in this limit are observables, the expectation values of field properties $ \langle f(\varphi) \rangle_{(\varphi|d)} $. Here, $ f(\varphi) = r^\dagger \varphi $ might be the integration of the field over some probing kernel $ r(x) $, or the like.  In case the field expectation of interest are robust with respect to further refinements of the pixelized field representation, the continuum limit of a field theory can be claimed to effectively be reached.

\section{Free fields}\label{sec:free}
\subsection{Simplistic scenario}\label{sec:simplescenario}

To have an instructive problem, let us discuss the simplest IFT scenario of a \textit{free field} first. Free means that the information Hamiltonian will be only be of quadratic order in the field, leading to a linear operation translating the data into field estimates, for which the superposition principle holds.

Imagine we observe at some time $t=0$ relative temperature fluctuations
\begin{equation}\label{eq:deltaT}
\varphi(x,t)=\frac{\delta T(x,t)}{\overline{T}} \ll 1
\end{equation}
of a temperature field $T(x,t)$ that is defined over one spatial and one temporal dimension, $ \Omega=\mathbb{R}^2 $, with the spatial and temporal mean temperature $\overline{T} = \langle T(x,t) \rangle_{(x,t)}$.
These fluctuations are excited by some external Gaussian excitations $ \xi \hookleftarrow \mathcal{G}(\xi,\Xi)$ and decay by diffusion:
\begin{equation}\label{eq:pde}
\partial_t \varphi(x,t) = \kappa\, \partial^2_x \varphi(x,t) + \nu^{1/2}\, \xi(x,t),
\end{equation}
with $ \kappa $ the spatial diffusion coefficient and $\nu$ being a rate%
\footnote{The external excitation fluctuations should add up under time integration in quadrature, and if $\xi$ is as dimensionless as $\varphi$, then $[\nu] = \mathrm{sec}^{-1}$.}.

Fourier transforming%
\footnote{We use the (one dimensional) Fourier convention $ \varphi_k = \int \mathrm{d}x\,e^{ikx} \varphi_x$,  $ \varphi_x = \int \mathrm{d}k/(2\pi)\,e^{-ikx} \varphi_k$. Therefore, the identity operator in Fourier space reads $ \mathbb{1}_{kk'} = 2\pi\,\delta(k-k')$. Furthermore, we use arguments and indices interchangeably and to denote whether a quantity is in real space or Fourier space: $ \varphi_x = \varphi(x) $ and $ \varphi_k = \varphi(k) =  \int \mathrm{d}x\,e^{ikx} \,\varphi(x)$.}
 this equation with respect to space and time yields
\begin{equation}
\varphi(k,\omega) = \frac{\nu^{1/2} \xi(k,\omega)}{i\,\omega+\kappa\, k^2}.
\end{equation}
For simplicity, we assume the excitation to be white and of unit variance,
$ \Xi_{(x,t)\,(x',t')}= \delta(x-x')\,\delta(t-t') $ or briefly $\Xi = \mathbb{1}$. This implies that in Fourier space the excitation is white as well, $ \Xi_{(k,\omega)\,(k',\omega')}= (2\pi)^2\,\delta(k-k')\,\delta(\omega-\omega') $.  The field covariance in Fourier space is then
\begin{eqnarray}\label{eq:Phi_omegak}
\Phi_{(k,\omega)\,(k',\omega')}  &=& \langle \varphi_{(k,\omega)} \, \varphi^*_{(k',\omega')} \rangle_{(\varphi)} \nonumber\\
&=&
 \frac{(2\pi)^2\,\nu}{\omega^2+\kappa^2 k^4}\,\delta(k-k')\,\delta(\omega-\omega').
\end{eqnarray}

Assume at time $ t=0 $ we obtain some data on the field according to a linear measurement as described by Eqs.~\ref{eq:d} and \ref*{eq:d_comp}, with Gaussian, field-independent measurement noise $ n \hookleftarrow \mathcal{G}(n,N)$ that is white ($ N=\sigma^2_n\,\mathbb{1} $) and has variance $\sigma^2_n$. Thus, the likelihood is
\begin{equation}
\mathcal{P}(d|\varphi) = \mathcal{G}(d-R\,\varphi,N).
\end{equation}

To simplify the discussion further, we assume that the measurement probes every location, such that $r_i(x) = \delta(i-x)$ or put differently $ R=\mathbb{1} $. This is an idealization, as now an infinite number of locations are probed. Note, that this still does not set us into a perfect knowledge about the field, as the data vector $ d=\varphi +n $ is still corrupted by noise.

To infer the field configuration at $ t=0 $ via Bayes' theorem, Eq.~\ref{eq:Bayes_theorem}, we need the prior. As the dynamics is linear and the excitation is Gaussian, the field statistics will be Gaussian, $\mathcal{P}(\varphi) =\mathcal{G}(\varphi,\Phi) $. This requires that we work out the field covariance for a specific time. Fourier back-transforming Eq.~\ref*{eq:Phi_omegak} with respect to time gives
\begin{equation}\label{eq:Phi_tk}
\Phi_{(k,t)\,(k',t')}  = \frac{\pi\,\nu}{\kappa\,k^2}\,\delta(k-k')\,\exp(-|t-t'|\,\kappa\,k^2).
\end{equation}
Temperature fluctuations decay in time and they do this faster as smaller their spatial scale. Here, we are only interested in the spatial correlation structure, since we restrict our analysis to a single time slice at $t=0$ in order to keep the discussion simple.

The equal time correlation is
\begin{equation}\label{eq:Phi_0k}
\Phi_{(k,t)\,(k',t)}  = \frac{\pi\,\nu}{\kappa\,k^2}\,\delta(k-k') = 2 \pi\,\delta(k-k')\,  P_\varphi (k),
\end{equation}
with 
\begin{equation}
P_\varphi (k)= \frac{\nu}{2\,\kappa\,k^2}
\end{equation}
the spatial power spectrum of the field.

From now on we only consider the spatial domain, $ \Omega=\mathbb{R} $. A field realization $\varphi$ of a Gaussian process with this covariance $\Phi$ is shown in Fig.~\ref{fig:Wiener_filter}.

\subsection{Prior regularization}\label{sec:regularization}
The covariance of our temperature fluctuation field is actually equivalent to regularization of the solutions by an $\mathcal{L}^2$-norm on the field gradient. The prior Hamiltonian,
\begin{eqnarray}
\mathcal{H}(\varphi) &=& -\ln \mathcal{P}(\varphi) = -\ln \mathcal{G}(\varphi, \Phi) \nonumber\\
&=&
\frac{1}{2} \varphi^\dagger \Phi^{-1} \varphi +  \frac{1}{2} \ln |2\pi\,\Phi|\nonumber\\
&=& \frac{1}{2} \int \frac{\mathrm{d}k}{2\pi}\,\frac{2\,\kappa}{\nu}\,k^2|\varphi_k|^2 + \mathrm{const}\nonumber\\
&=&
\frac{\kappa}{\nu} \int \mathrm{d}x\, |\nabla\, \varphi|^2 + \mathrm{const}\nonumber\\
&=&
\frac{\kappa}{\nu} ||\nabla\, \varphi||^2 + \mathrm{const},\label{eq:H(phi)}
\end{eqnarray}
is regularizing the joint information Hamiltonian of data and field,
\begin{eqnarray}
\mathcal{H}(d,\varphi)&=&  \mathcal{H}(d|\varphi) + \mathcal{H}(\varphi),\label{eq:H(d,phi)}
\end{eqnarray}
with $  \mathcal{H}(d|\varphi)= -\ln \mathcal{P}(d|\varphi) $ the likelihood information.

Since we started with a physical model, the strength of this regularization is specified by the ratio of the diffusion constant to the excitation rate $ \kappa/\nu $, a physical quantity. This should be seen in contrast to the frequently used ad-hoc parameter put in front of an $\mathcal{L}^2$-gradient regularization term $||\nabla\, \varphi||^2 \equiv \int \mathrm{d}x\, |\nabla\, \varphi|^2$.
Physics herself provides the necessary regularization for us to treat the otherwise ill-posed problem. No ad-hoceries are needed.

\subsection{Wiener filter}

Now, we can work out the information Hamiltonian
\begin{eqnarray}
\mathcal{H}(d,\varphi) 
&=& \frac{1}{2} \varphi^\dagger \Phi^{-1} \varphi +
    \frac{1}{2} (d-R\,\varphi)^\dagger N^{-1} (d-R\,\varphi)\nonumber\\
&&    +  \frac{1}{2} \ln |2\pi\,\Phi|+  \frac{1}{2} \ln |2\pi\,N|\label{eq:H(d,phi)2}
\end{eqnarray}
as well as the logarithmic partition function
\begin{eqnarray}
\ln \mathcal{Z}(d,J)&=& \ln \int \mathcal{D}\varphi \, e^{-\mathcal{H}(d,\varphi)+J^\dagger \varphi} \nonumber\\
&=& \frac{1}{2}\, (j+J)^\dagger D\, (j+J) - \frac{1}{2}\, d^\dagger N^{-1} d  \nonumber\\
&& +\frac{1}{2}\, \ln |2\pi\,D|-\frac{1}{2}\, \ln |2\pi\,\Phi|- \, \frac{1}{2} \ln |2\pi\,N|
\label{lnZ}
\\
\mbox{with }D&=& \left( \Phi^{-1} + R^\dagger N^{-1} R\right)^{-1}\label{D}\\
\mbox{and }j &=& R^\dagger  N^{-1}d\label{j}.
\end{eqnarray}
It turns out that a separate moment generating variable $J$ is not needed, as the variable $j$ can take over this role. We just set $J=0$, take derivatives with respect to $j$, but do not set it to zero afterwards. The first and second connected moments of the posterior field are then
\begin{eqnarray}
 \langle \varphi \rangle_{(\varphi|d)}^{(\mathrm{c})} &=&
 \frac{\partial \ln \mathcal{Z}(d,j) }{\partial j^\dagger }
 = D\, j = m  \label{m}\label{eq:m=Dj}\\
  \langle \varphi\,\varphi^\dagger  \rangle_{(\varphi|d)}^{(\mathrm{c})}  &=&
\frac{\partial^2 \ln \mathcal{Z}(d,j) }{\partial j^\dagger \partial j  } = D \label{eq:D}.
\end{eqnarray}

\begin{figure}
	\centering
	\includegraphics[width=\linewidth, trim={1cm 0.45cm 0.8cm 0},clip]{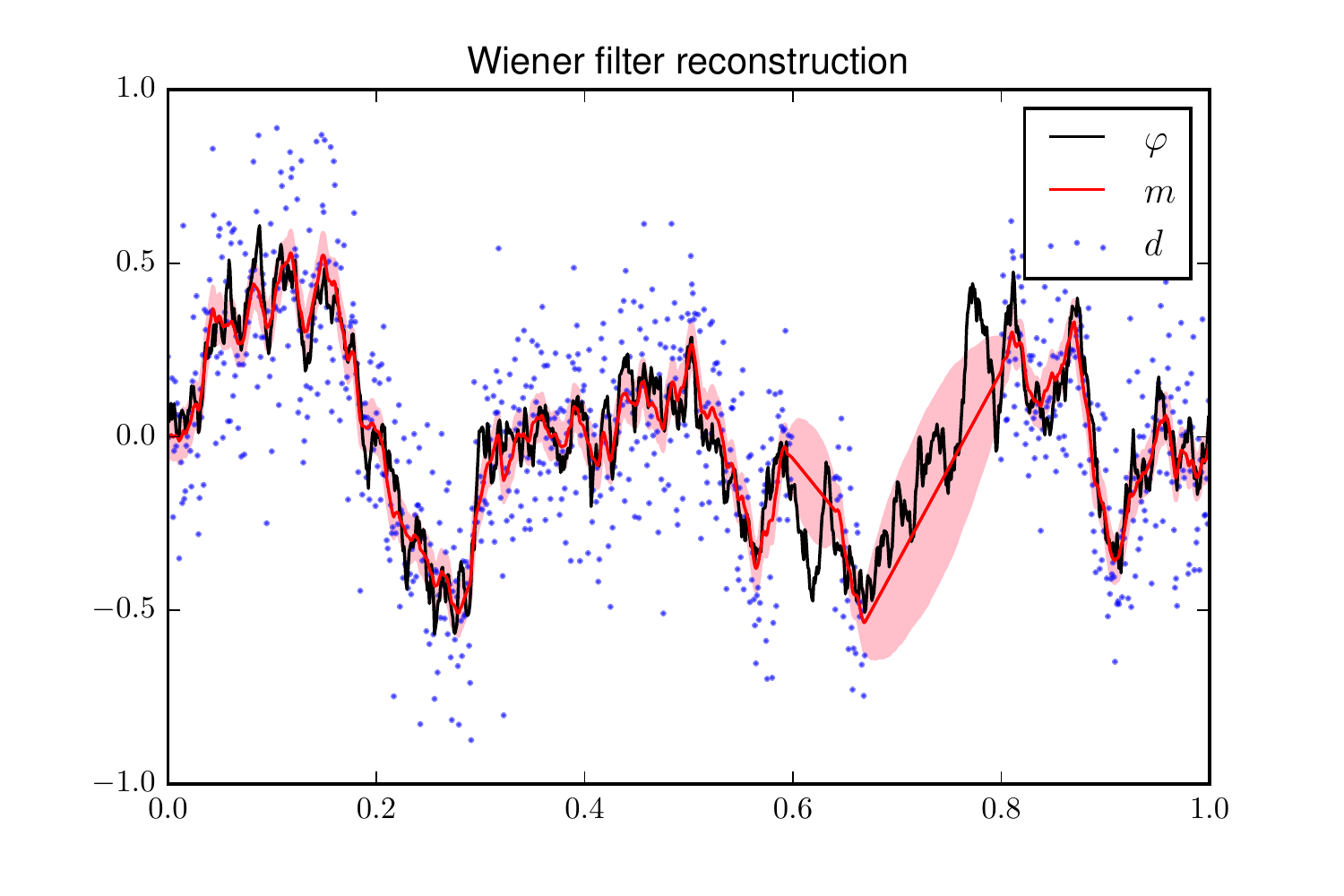}
	\caption{Wiener filter reconstruction $m$ (Eq.~\ref{eq:WF}) of the unknown field $\varphi$ from noisy and incomplete data $d$ as discussed in the text. The 1-$\sigma$ posterior uncertainty is shown as well (Eqs.~\ref{eq:sigma} and \ref{eq:DWF}). Note the enlarged uncertainty in regions without data. The maximum likelihood field estimate is identical to the data points.}
	\label{fig:Wiener_filter}
\end{figure}

Higher order connected moments vanish. This means that the posterior is a Gaussian with this mean and covariance, 
\begin{equation}
	\mathcal{P}(\varphi|d) = \mathcal{G}(\varphi -m, D),\label{eq:Gauss2}
\end{equation}
an insight also a direct calculation based on rewriting Eq.~\ref{eq:H(d,phi)2} in terms of these moments verifies:
\begin{eqnarray}
\mathcal{H}(d,\varphi) &\widehat{=}& \frac{1}{2} \varphi^\dagger D^{-1} \, \varphi + \frac{1}{2}\left( \varphi^\dagger j + j^\dagger \varphi \right) \nonumber\\
&=& \frac{1}{2} \varphi^\dagger D^{-1} \, \varphi + \frac{1}{2}\left( \varphi^\dagger D^{-1} \underbrace{D\, j}_{=m} + j^\dagger D\, D^{-1} \varphi  \right) \nonumber\\
&\widehat{=}& \frac{1}{2}  \left( \varphi-m\right)^\dagger D^{-1} \left( \varphi-m\right) \label{eq:completion}\\
\Rightarrow \mathcal{P}(d,\varphi) &\propto& e^{-\mathcal{H}(d,\varphi)} \propto  \mathcal{G}(\varphi -m, D).\label{eq:propto}
\end{eqnarray}
In Eq.~\ref{eq:completion}, we collected the quadratic and linear terms in $\varphi$, introduced $\widehat{=}$ to express equality up to irrelevant constants, used that $D$ is invertible and self-adjoint, and performed a quadratic completion. The proportionalities in Eq.~\ref{eq:propto} combine into an equality, since the probabilities on the left and right hand sides of the equations are both properly normalized. Hence, the posterior is a Gaussian, with mean $m=D\,j$ and uncertainty covariance $D$.

\begin{figure}
	\centering
	\includegraphics[width=\linewidth, trim={1cm 0.45cm 0.8cm 0},clip]{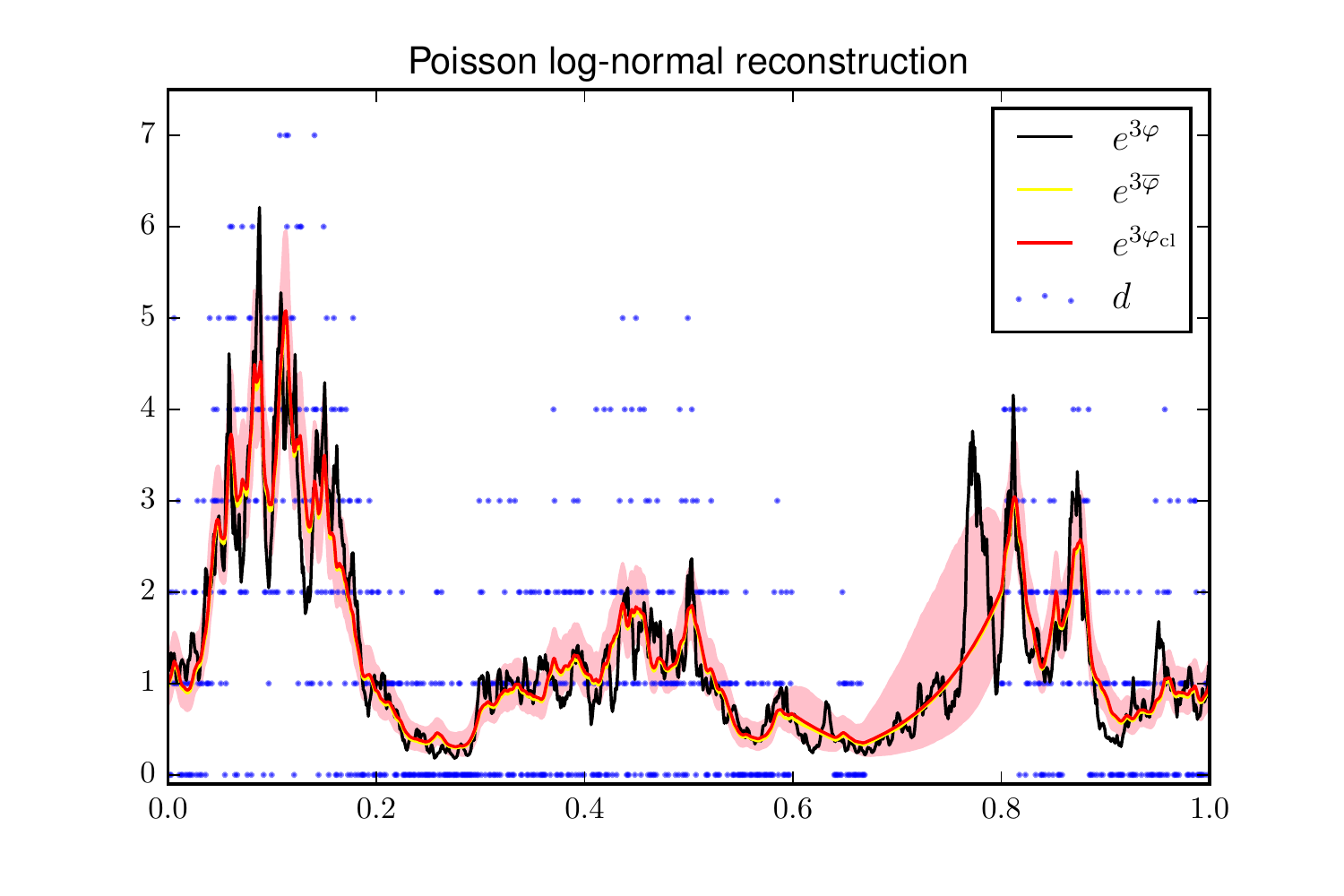}
	\caption{Classical field reconstruction $\varphi_\mathrm{cl}$ (Eq.~\ref{eq:MAP}) of the unknown field $\varphi$ from Poisson counts $d$ subject to a non-linear, exponential response. The 1-$\sigma$ posterior uncertainty is shown as well  (Eqs.~\ref{eq:sigma} and \ref{D_MAP}). Note the enlarged uncertainty in regions without data. The field realization $\varphi$ is the same as in Fig.~\ref{fig:Wiener_filter}. The ML field  $e^{3\varphi_\mathrm{ML}}=d$ and the posterior mean field $e^{3\overline{\varphi}} \approx e^{3\varphi_\mathrm{HMC}}$ are shown as well. The HMC results were calculated by the HMCF code  \cite{2018arXiv180702709L} and kindly provided by its author, Christoph Lienhard.}
	\label{fig:Poisson}
\end{figure}

For our specific problem we have $ R=\mathbb{1} $, $ N=\sigma_n^2 \, \mathbb{1} $, and
\begin{equation}
	\Phi_{kk'}  =\frac{2\pi \,\delta(k-k')}{2\,\kappa\, \nu^{-1} k^2}
\end{equation}
and therefore
\begin{eqnarray}
D_{kk'} &=& 2\pi \,\delta(k-k')\,\left[ {2\,\kappa\, \nu^{-1} k^2} + \sigma_n^{-2} \right]^{-1},\\
j_k     &=& \sigma_n^{-2} d_k,\mbox{ and} \\
m_k     &=& \int \frac{\mathrm{d}k'}{2\pi} D_{kk'} j_{k'} =
		    \frac{d_k}{1+{\lambda^2 k^2}}\mbox{ with } \\
\lambda^2 &=& \frac{2\,\kappa}{\nu}\,  \sigma_n^{2}.
\end{eqnarray}

This means that the optimal field reconstruction $ m $ follows the data on large spatial scales $  1/k \gg  \lambda$ (as $m_k \approx d_k$ there) and should be a strongly suppressed version of the (Fourier space) data for small spatial scales  with $ 1/k \ll  \lambda$ (as $m_k \approx d_k\,(\lambda\,k)^{-2} \ll d_k$ there). The optimal reconstruction is the result of a Fourier space filter operation applied to the data. This result was first found by Wiener \cite{1949wiener-short} and the resulting filter is therefore called the \textit{Wiener filter}.

It should be noted that in ad-hoc regularization, the parameter $ \lambda $, which controls the threshold wavelength of the low-pass filter, is chosen out of the blue, whereas here the knowledge about the statistics of the signal is determined by physical quantities like $\lambda = \sqrt{2\,\kappa/\nu}\,\sigma_n$.

\begin{figure}
	\centering
	\includegraphics[width=\linewidth, trim={1cm 0.45cm 0.8cm 0},clip]{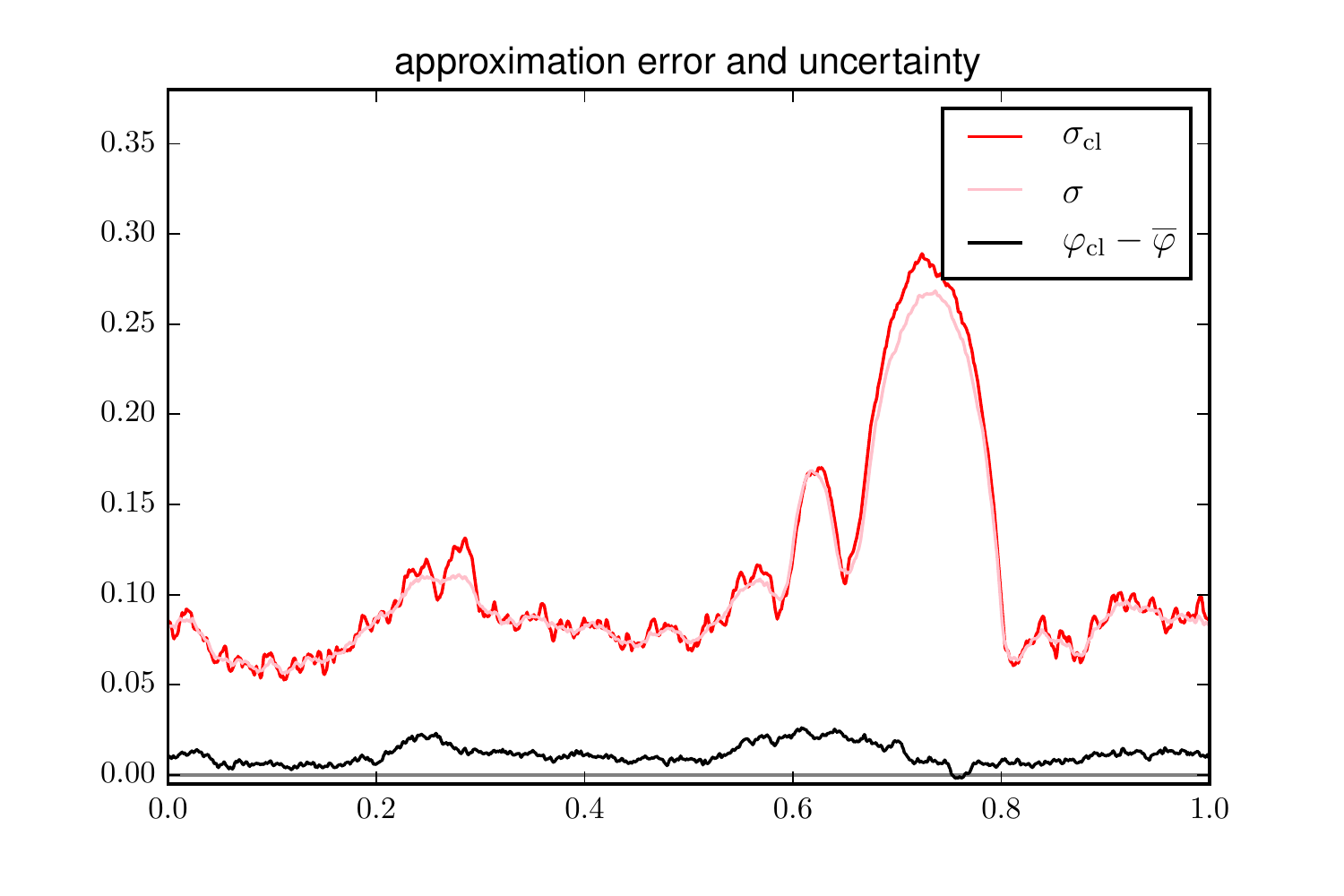}
	\caption{Approximation errors of the MAP field estimate $\varphi_\mathrm{cl}$ and its uncertainty $\sigma_\mathrm{cl}$ displayed in Fig.\ref{fig:Poisson} with respect to their more accurate MHCF estimates $\overline{\varphi}$  and $\sigma$, respectively. Black: Absolute error of the MAP estimate revealing a small systematic overestimation of it. Red: Uncertainty via Laplace approximation around MAP estimate. Pink: Uncertainty of the HMC estimate. Note the enlarged uncertainties at locations without data or fewer counts.}
	\label{fig:errorPoisson}
\end{figure}

\subsection{Generatlized Wiener filter}

We made a number of assumptions in this derivation, namely that the noise is white, the field has a specific covariance, and the instrument response is the identity. None of these were essential. Assuming colored noise with covariance $N$, a general field covariance $\Phi$, and an arbitrary (but linear) response $R$ would have provided us with the solution given by Eqs.~\ref{D} -
\ref{eq:Gauss2}, Solely the nice property of this generalized Wiener filter being a Fourier-space-only operation would have been lost. 

The price for this \textit{generalized Wiener filter} is that applying $D$ to $j$ to obtain the posterior mean $m=D\,j$ can then often not be done analytically anymore. Instead, one needs to use a numerical scheme like the conjugate gradient method to solve 
\begin{equation}
D^{-1} \, m = j
\end{equation}
for $m$. This is possible in case $D^{-1}= \Phi^{-1} + R^\dagger N^{-1} R$ is available as an operator, a computer routine that takes and returns a vector in signal space. This requires only that  its  constituents  $\Phi^{-1}$,  $ N^{-1}$,  $R$, and  $R^\dagger$ are available as computer routines as well.
The implementation of the translational invariant inverse prior field covariance operator  $\Phi^{-1}$ is done via a fast Fourier transformation of the field, a division with the power spectrum, followed by a back transformation. This way, by the usage of an operator representation and the conjugate gradient method, very complex measuremnt situations can be handled efficiently.  One such situation is shown in Fig.~\ref{fig:Wiener_filter}.

\subsection{Information propagation}

Before we treat more complicated measurement situations, let us try to develop some intuition what the (generalized) Wiener filter does. It turns linear data $ d=R\,\varphi + n  $ on a field into an optimal reconstruction $ m $ (see Fig.~\ref{fig:Wiener_filter}) with
\begin{eqnarray}
	m &=& D\,j,\label{eq:WF}\\
	j &=& R^\dagger N^{-1}d,\mbox{ and}\\
	D &=& \left( \Phi^{-1} + R^\dagger N^{-1} R\right)^{-1}.\label{eq:DWF}
\end{eqnarray}
First, we inspect $j$. This is the inversely noise weighted data $ N^{-1}d $ back projected\footnote{The technical term \textit{back projection} is to be understood in the way light is projected from a source (here the data living in data space) onto a screen (here the signal space) and not in the way it is used in the mathematical literature, where projection operators are allways endomorphic, staying within the same space.} by the adjoint response $R^\dagger$ into the field domain $\Omega$. Every piece of data is thrown onto all  locations that have influenced it in the measurement process, with exactly the strength of this influence, as encoded in the adjoint (or transposed) response $ R^\dagger $. Since $ j $ carries the essential information of the data and sources our posterior knowledge on the field, the name \textit{information source} seems to be appropriate.

Now, let's see how $ D $ acts on $ j $ in $ m=D\,j $. In components, this is
\begin{equation}
m(x) = \int_{\Omega } \!\!\mathrm{d}x \, D(x,y)\,j(y).
\end{equation}
Thus, $ D(x,y) $ transports (and weights) the information source $j$ from all locations $ y $ to location $ x $, where they are added up to provide the posterior mean field. For this reason  the term \textit{information propagator} is natural for $ D $. This nomenclature is very much in accordance with conventions in quantum field theory and also serves IFT well.

For our simplistic example, the information propagator in position space
\begin{equation}\label{D_example}
D(x,y) = \frac{\sigma_n^2}{2\lambda}e^{-\frac{|x-y|}{\lambda}}
\end{equation}
is a translation-invariant convolution kernel, since neither the prior nor measurement singles out any location. The mean field $ m $ most strongly perceives the local information source $ j_x = d_x/\sigma_n^2 $ at every location $ x $, but also more distant contributions with exponentially decaying weights:
\begin{equation}
m(x) = \int  \frac{\mathrm{d}y}{2\lambda}e^{-\frac{|x-y|}{\lambda}}\,d(y).
\end{equation}
This operation averages down the noise in an optimal fashion.

The information propagator $D$ is informed about how the measurement was performed as it incorporates $ R $ and $ N $. Furthermore, $D$ knows about the correlations of field values at different locations through $ \Phi $. It uses the first to properly weight the information source and the latter to extrapolate it to locations not or only poorly probed by the instrument.

Furthermore, since $D$ is also the posterior uncertainty covariance in $\mathcal{P}(\varphi|d)=\mathcal{G}(\varphi -m, D) $, it also encodes how certain we can be about the reconstructed field via
\begin{equation}
\sigma_{\varphi_x}^2 \equiv \langle (\varphi_x-m_x)^2 \rangle_{(\varphi_x|d)} = D_{xx}\label{eq:sigma}
\end{equation}
and how much the uncertainty is correlated between different locations (via $D_{xy}$). Eq.~\ref{D_example} illustrates that in case of larger noise (as controlled by $ \sigma_n $) this uncertainty is larger, since
$\sigma_{\varphi_x}^2= D_{xx} = \sigma_n^2/(2\,\lambda) = \sigma_n/\sqrt{8\,\kappa/\nu}\propto \sigma_n$. It is correlated over larger distances, since $ \lambda = \sqrt{2\kappa/\nu}\,\sigma_n \propto \sigma_n$.  Fig.~\ref{fig:Wiener_filter} also shows the $\sigma_{\varphi_x}$-uncertainty range around the general Wiener filter reconstruction $m$.

\subsection{Maximum likelihood}

It is instructive to see how much the field reconstruction using the Wiener filter benefitts from the knowledge of the correct correlation structure. We can remove this information by dropping the prior term $\mathcal{H}(\varphi)$ in Eqs.~\ref{eq:H(phi)}-\ref{eq:H(d,phi)2} and the subsequent calculation. The most conventient way to do so is to take the limit $\Phi^{-1} \rightarrow 0$, which yields for the field estimate
\begin{eqnarray}
m&=&D\,j= \left(\Phi^{-1} + R^\dagger N^{-1} R \right)^{-1} R^\dagger N^{-1} d \nonumber\\
&\rightarrow & 
\left(R^\dagger N^{-1} R \right)^{-1} R^\dagger N^{-1} d\nonumber\\
&= & 
\left(\mathbb{1}^\dagger \left(\sigma_n^{2} \mathbb{1} \right)^{-1} \mathbb{1} \right)^{-1} \mathbb{1}^\dagger \left(\sigma_n^{2} \mathbb{1} \right)^{-1} d\nonumber\\
&= & d.
\end{eqnarray}
Thus, by ignoring the prior knowledge on the field correlation one is forced in this case to take the data as the best field estimate. Fig.~\ref{fig:Wiener_filter} shows that this can be a very poor guess.

For the posterior, the mean, the median, and the mode of the distribution coincide, as it is a Gaussian. Removing the prior information therefore has transformed our field estimate into a maximum likelihood (ML) estimate. ML is a popular statistical practice that is known to over-fit the data. In our case, the ML estimation absorbed the full noise into the solution, due to the lack of any redundancy in the data and the absence of the regularizing influence of the prior Hamiltonian.

\subsection{Optimal estimate}

As the ML field estimate performed so poorly, the question arises, whether a better estimate than the Wiener filter solution exists. To answer this question, a definition of \textit{better} or of \textit{worse} is required.

With the reconstruction error field $\varepsilon(x) = \psi(x) - \varphi(x)$ of the reconstruction $\psi $  of a field $\varphi$ a loss function $l(\varepsilon)$ can be defined. Averaging the loss over the posterior provides the expected loss of a chosen $\psi$, which can be regarded as an objective function
\begin{equation}
L(\psi|d) =  \langle l(\psi-\varphi) \rangle_{(\varphi|d)}=
\int\,\mathcal{D}\varphi \,\mathcal{P}(\varphi|d) \,l(\psi-\varphi)
\end{equation}
for the field estimation. Minimizing $ L(\psi|d)$ with respect to $\psi$ then provides an optimal field estimate according to the chosen error metric.

Popular loss functions are the square error norm $l(\varepsilon)= || \varepsilon||^2 = \int dx\, \varepsilon^2(x)$, the absolute error norm 
$l(\varepsilon)= | \varepsilon|_1 =\int dx\, |\varepsilon(x)|$, and the
delta-loss $l(\varepsilon)=-\delta(\varepsilon)$. These lead to the posterior mean, median, and mode as the optimal estimators, respectively. For the posterior mean being the optimal estimator under an $\mathcal{L}^2$-error norm, the proof had already be given in Eq.~\ref{eq:L2mean}.

For the Gaussian posterior, we encountered so far, mean, median, and mode coincide, as does any estimator based on a monotonically increasing error loss. Thus, we can state that the Wiener filter reconstruction $m$ is the optimal estimator with respect to any sensible error metric. No better estimate exists in this case. 

\subsection{Critical filter}\label{sec:critical}

So far, the prior signal covariance $\Phi$ was assumed to be known. If it is unknown, it has to be inferred as well. Here, we will sketch a simple \textit{empirical Bayes} method for infering both, the signal and its spectrum, the \textit{critical filter}.

In empirical Bayes, the joint estimate of two interrelated quantities $a$ and $b$ is performed by using a point estimate of one quantity (\textit{e.g.} $b=b_\mathrm{MAP})$ in the estimation of the other. The logic is captured in the following calculation:
\begin{eqnarray}
\langle a \rangle_{(a,b|d)} &=& \int da\,\int db\, a\,\mathcal{P}(a,b|d)\nonumber\\
&=& \int db\, \mathcal{P}(b|d) \int da\, a\,\mathcal{P}(a |b, d) \nonumber\\
&\approx&  \int db\, \delta[b-b_\mathrm{MAP}(d)]\, \mathcal{P}(a |b, d) \int da\, a\,\mathcal{P}(a |b, d) \nonumber\\
&=&  \langle a \rangle_{(a|b_\mathrm{MAP}(d),d)} .
\end{eqnarray}
Thus, the uncertainty in $\mathcal{P}(b|d)$, the $a$-marginalized posterior for $b$, is ignored. In our specific case, $a=\varphi$ will be the signal field and $b=\Phi$ its covariance. We recall that $\mathcal{P}(\varphi|\Phi) = \mathcal{G}(\varphi, \Phi)$.

A hyperprior, a prior on a prior quantity, is necessary for this. In case statistical homogeneity can be assumed,  $\Phi$  is diagonal in Fourier space,  $\Phi_{\vec{k}\vec{q}}= (2\pi)^u \,\delta^u(\vec{k}-\vec{q})\,P_\varphi(\vec{k})$, with  $ P_\varphi(\vec{k})$ the spectral power density and $\vec{k}$ a Fourier space vector, denoted here in bold face to distinguish it from its length  $k = |\vec{k}|$. In case statistical isotropy can be assumed, the spectral power density is isotropic $P_\varphi(\vec{k})=  P_\varphi(k)$, and all Fourier modes with the same length $k = |\vec{k}|$ have the same power.

It is then convenient to introduce the spectral band projector
$(\mathbb{P}_k)_{\vec{q}\vec{q}'}\equiv
(2\pi)^u\delta(\vec{q}-\vec{q}')\,
\delta(|\vec{q}|-k)$. With this the maximum a posteriori estimator for the power spectrum\footnote{For this $\mathcal{P}(P_{\varphi}|d)= \int \mathcal{D}\varphi\, \mathcal{P}(P_{\varphi},\,\varphi|d) $ has to be maximized, the power spectrum posterior, which is marginalized over the field.}
is
\begin{equation}\label{eq:critical}
P_{\varphi}(k)=\frac{\mathrm{Tr}\left[\left(m\,m^{\dagger}+D\right)\,\mathbb{P}_{k}\right]}{\mathrm{Tr}\left[\mathbb{P}_{k}\right]},
\end{equation}
where a flat prior on the spectral power density entries was assumed \cite{2011PhRvD..83j5014E}. A compact derivation of this result can be found on Wikipedia \cite{wiki:xxx}.

Eq.~\ref*{eq:critical} and \ref{eq:m=Dj} have to be solved jointly, as $m=D\,j$ depends on the power spectrum $P_{\varphi}(k)$ through $D$, and the power spectrum itself also depends on  $m$. This so-called \textit{critical filter} therefore iteratively estimates a field and the power spectrum of the process that has generated it from data. 

The field reconstruction is regularized by the covariance structure implied by the power spectrum, and the spectrum reconstruction is informed by the field estimate. Thus, the data is used in a non-linear way to reach the final critical filter field estimate and we have an example of an \textit{interacting field theory}, where the field solution depends in a non-linear way on the data.

The critical filter as presented above performes well, in case there are sufficiently many Fourier modes informing $P_\varphi(k)$. Otherwise, additional prior information on the spectrum should be used. For example, the usage of a spectral smoothness prior \cite{2013PhRvE..87c2136O} has turned out to be very useful in cases where such an assumption is justified.

The critical filter, its variations and extensions\footnote{For informative spectral prior  \cite{2011PhRvD..83j5014E}, spectral smoothness hyper-priors \cite{2013PhRvE..87c2136O}, unknown noise \cite{2011PhRvE..84d1118O}, non-linear measurement \cite{2013PhRvE..87c2136O, 2017arXiv171102955K} and others.} play an important role for real world applications of IFT, since very often a sufficiently well understood theory that predicts the statistical properties of a field is not available. With the critical filter, a field and its power spectrum can be estimated simultaneously from the same data.

\section{Interacting fields}\label{sec:interactingIFT}

The generalized Wiener filter of the free IFT appeared under a number of assumptions. These were Gaussianity and mutual independence of signal and noise, linearity of the measurement and the precise knowledge of all involved operators ($ \Phi $, $ R $, and $ N $).

In case these conditions are not given, the situation becomes more complicated. This is expressed by the information Hamiltonian $ \mathcal{H}(d,\varphi)=-\ln\mathcal{P}(d,\varphi) $ getting terms of higher than second order in the field $ \varphi $ (after marginalization of all other unknowns), so-called \textit{interaction terms}. These lead -- in general -- to non-linear relations between the data and mean field as well as to non-Gaussian posteriors. The theory becomes non-free or interacting.

Let us stay specific and turn our simplistic example from above into a non-linear problem. To do so we modify Eq.~\ref{eq:deltaT} that it can cope with large temperature fluctuations without becoming unphysical ($ = $ allowing negative temperatures). We achieve this by setting now
\begin{equation}
\varphi(x) = \ln\left(\frac{T(x)}{\overline{T}}\right)
\end{equation}
such that
\begin{equation}
T(x) = \overline{T}\,\exp(\varphi(x))
\end{equation}
is now a strictly positive temperature, irrespective of the value of $ \varphi(x)$, which we still assume to be drawn from a zero centered Gaussian, $ \varphi \hookleftarrow \mathcal{G}(\varphi,\Phi) $, with covariance given by Eq.~\ref{eq:Phi_0k}.

Let us assume that we are able to observe the thermal photons emitted from our system, such that the expected number of received photons by a detector in an observational period  is
\begin{eqnarray}
\mu_i(\varphi) &=& \int \mathrm{d}x \,r'_i(x)\, T^3(x) = \int \mathrm{d}x \,r_i(x)\, \exp[3\,\varphi(x)]  \nonumber\\
&=&  r_i^\dagger e^{3\varphi},
\end{eqnarray}
where with $ i $ we label the detector and the observational period, we have absorbed physical constants into the responses $ r' $, have set $ r = r'\,\overline{T}^3$, and introduced the convention to apply functions point-wise to fields, $(e^{\varphi})_x=e^{\varphi_x}$.

Observing photons is a Poisson process, and each data bin has an independent distribution function
\begin{equation}
\mathcal{P}(d_i|\mu_i) = \frac{\mu_i^{d_i}}{d_i!}\,e^{\mu_i}
\end{equation}
for the number of photons $ d_i $ received by the individual detectors' observations as  labeled by $i$ . For the full dataset we write $ \mu = R\,\exp(3\varphi) $, where we define the linear part of the response operator as $ R_{ix}=r_{i}(x) $.

Thus, all elements to construct the information Hamiltonian are available, and we get
\begin{eqnarray}
\mathcal{H}(d,\varphi) &=& \mathcal{H}(\varphi) + \sum_i \mathcal{H}(d_i|\mu_i(\varphi))\nonumber\\
&\widehat{=}& \frac{1}{2}\,\varphi^\dagger \Phi^{-1} \varphi + 1^\dagger R\,e^{3\varphi} - d^\dagger \ln \left( R\,e^{3\varphi} \right),\label{eq:HLNP}
\end{eqnarray}
where $ \widehat{=} $ denotes that we dropped irrelevant (since $ \varphi $-independent) constants and $ 1^\dagger  $ is a row vector in data space with all entries being 1. This clearly is an information Hamiltonian with strong self-interaction terms that are provided by the an-harmonic orders of the exponential function and the logarithm.

To treat this, we will use two approximative approaches from theoretical physics, by calculating the classical and the mean field solution.

\subsection{Classical field approximation}

The classical field $\varphi_\mathrm{cl}$ is simply given by the minimum of the Hamiltonian:
\begin{eqnarray}
	0&=& \frac{\partial \mathcal{H}(d,\varphi)}{\partial \varphi}
=	 \Phi^{-1} \varphi +  e^{3\varphi} \,R^\dagger 3 -  \frac{e^{3\varphi} \,R^\dagger (3\,d)}{ R\,e^{3\varphi}} \label{eq:HLNPgrad}\\
	 &=&
	 \Phi^{-1} \varphi +  3\, e^{3\varphi} \,R^\dagger \left( 1- \frac{d}{ R\,e^{3\varphi}}\right) \Rightarrow \nonumber\\
	 \varphi_\mathrm{cl} &=&    3\,\Phi\, e^{3\varphi_\mathrm{cl}} \,R^\dagger \left(  \frac{d- R\,e^{3\varphi_\mathrm{cl}}}{ R\,e^{3\varphi_\mathrm{cl}}}\right)\label{eq:MAP}
\end{eqnarray}
A solution to this non-linear integral equation should exhibit $d \approx R\,e^{3\varphi_\mathrm{cl}}$ as this satisfies the likelihood, but will also be smoothed due to the application of the smoothing action of the covariance kernel $ \Phi $.

The classical field minimizes the Hamiltonian and, according to Eq.~\ref{eq:BayesH}, maximizes the posterior. Thus, the classical field is the \textit{maximum a posteriori} (MAP) field estimate. In general the classical or MAP solution is not identical to the posterior mean. This requires special conditions, like that the posterior is point symmetric about $\varphi_\mathrm{cl}$, as it is the case of the free theory discussed in Sect.~\ref{sec:free}.

An estimate of the uncertainty covariance is given by the inverse Hessian of the Hamiltonian at its minimum\footnote{The middle term of the following exppression is small at the minimum, as there $R(\varphi)\approx d$. Since this term can produce negative eigenvalues of the Hessian, prohibiting its usage as a covariance, it is often (and also here) dropped in numerical applications of IFT.}
\begin{eqnarray}
({D}^{-1})_{xy} &\approx&
\left(\frac{\partial^2 \mathcal{H}(d,\, \varphi_\mathrm{cl}) }{\partial \varphi_\mathrm{cl} \partial \varphi_\mathrm{cl}^\dagger}\right)^{-1}
\nonumber\\
&=&
\Phi^{-1}_{xy}
+ 9\,\delta_{xy} e^{3\,{\varphi_\mathrm{cl}}_x} \,R_x^\dagger
\left( 1- \frac{d}{ R\,e^{3\,\varphi_\mathrm{cl}}}\right) \nonumber\\
&&
+ 9 \, e^{3\,{\varphi_\mathrm{cl}}_x +3\,{\varphi_\mathrm{cl}}_y} \sum_i \,
\frac{R_{ix} \, R_{iy}\,d_i}{ (R\,e^{3\, \varphi_\mathrm{cl}} )_i^2}. \label{D_MAP}
\end{eqnarray}
The uncertainty in this Laplace approximation (treating Posterior as a Gaussian by expanding it around its maximum) clearly depends on the classical field, and therefore on the data. For a different data realization, the uncertainty would differ. Both, a field estimate and its uncertainty for this measurment situation, can be found in Fig.~\ref{fig:Poisson}.

\subsection{Comparison to other estimators}
\subsubsection{Maximum likelihood}

The MAP estimator provides the most probable field. A comparison to the ML field reveals how much this benefited from the prior information. Again, we remove the prior in Eqs.~\ref{eq:HLNP} and \ref{eq:HLNPgrad} by the limit $\Phi^{-1} \rightarrow 0$ to get from Eq.~\ref{eq:HLNPgrad} 
\begin{equation}
	d=R\,e^{3\,\phi_\mathrm{ML}}=e^{3\,\phi_\mathrm{ML}}
\end{equation}
Thus, the ML estimate of the photons emissivity $\varrho_\mathrm{ML} = e^{3\,\phi_\mathrm{ML}}$ is again given by the data. Fig.~\ref{fig:Poisson} shows how poorly the ML estimate compares to he MAP estimate. 

\subsubsection{Hamiltonian Monte Carlo for fields}

If the field estimate with the minimal square error is requested, the posterior mean should be calculated. This is not possible analytically anymore. Numerical methods can help here. One powerful method is \textit{Hamiltonian Monte Carlo} (HMC) sampling \cite{1987PhLB..195..216D,2017arXiv170102434B}, in which posterior samples are drawn from which the requested field estimate $\varphi_\mathrm{HMC}$ can be calculated by simple averaging. 

For the HMC the information Hamiltonian is extended to a full dynamical Hamiltonian by the introduction of an auxiliary momentum field $p(x)$:
\begin{eqnarray}
\mathcal{H}(d,\varphi, p) &= & \mathcal{H}(p| d,\varphi) + \mathcal{H}(d,\varphi)\mbox{ with}\\
 \mathcal{H}(p| d,\varphi) &=& \frac{1}{2} p^\dagger M^{-1} p + \frac{1}{2} \ln |2\pi M| ,
\end{eqnarray}  
where $M$ is a positive definite mass matrix, which optimally is chosen close to the information propagator $D$. The HMC algorithm constructs a new sample of $\varphi$ from an old one in three steps:
\begin{itemize}
	\item Draw a new initial momentum from
	\begin{equation}
		\mathcal{P}(p|d,\varphi)=\mathcal{G}(p,M).
	\end{equation}
	\item Follow the Hamiltonian dynamics from the current location $(\varphi, p)$ by integrating
	\begin{eqnarray}
		\frac{d\varphi}{dt}&=& \frac{\partial \mathcal{H}(d,\varphi,p)}{\partial p}\\
		\frac{dp}{dt}&=& -\frac{\partial \mathcal{H}(d,\varphi,p)}{\partial \varphi}
	\end{eqnarray}
	sufficiently long to generate an uncorrelated new location $(\varphi',p')$.
	\item Accept this new sample with the Metropolis-Hasting acceptance probability 
	\begin{equation}
			P(\mbox{accept}|\varphi',p',\varphi,p) = \mathrm{min}\left(1,e^{-\Delta \mathcal{H}}\right)
	\end{equation}
	with $\Delta \mathcal{H}= \mathcal{H}(d,\varphi',p')-\mathcal{H}(d,\varphi,p)$.
\end{itemize}
This results in samples drawn from
\begin{equation}
	\mathcal{P}(\varphi,p|d) = \mathcal{P}(\varphi|d)\,\mathcal{G}(p,M),
\end{equation} 
the direct product of the field posterior and a Gaussian in momentum.
The momenta can just be ignored as this distribution factorizes. 
The energy conserving property of the Hamilton dynamics integration ensures that only numerical errors lead to a small difference in the initial and final Hamiltonian and therefore the proposed samples  $(\varphi',p')$ are accepted with a high probability. The phase space volume conservation of the symplectic Hamiltonian dynamics ensures that probability mass is treated correctly. An appropriate symplectic numerical integration scheme needs to be adopted for this holding in the actual calculations as well. Thus, the HMC draws samples from the true posterior, and those are (largely) independent from each other.
 
An HMC implementation for fields (HMCF)\footnote{\url{https://gitlab.mpcdf.mpg.de/ift/HMCF}}  \cite{2018arXiv180702709L} was run on Eqs.~\ref{eq:HLNP} and the resulting posterior mean field is displayed in Fig.~\ref{fig:Poisson} as well. It is apparent that here the MAP field and the posterior mean agree very well, despite a small systematic difference on the 0.1-$\sigma$ level (see Fig.~\ref{fig:errorPoisson}). 

One could try to improve the MAP solution by replacing it with the so-called mean field approximation. In this particular case, this is not necessary, given the accuracy of the MAP solution here, however, in case of posterior distributions with long and asymmetric tails, MAP provides inferior estimates compared to the mean field approximation introduced next.

\subsection{Mean field approximation}

The mean field approximation $ m' $ should not be confused with the a posteriori mean field $ m \equiv \langle \varphi \rangle_{(\varphi|d)}$, which it tries to approximate. Mean fields in statistical field theory are constructed by minimization of the Gibbs free energy,
\begin{equation}
G(m',D')= U(m',D') - T\,S(D').
\end{equation}
In IFT the internal energy $U$ is defined as
\begin{equation}
U(m',D') = \langle \mathcal{H}(d,\varphi) \rangle_{\mathcal{G}(\varphi-m',D')},
\end{equation}
the information Hamiltonian averaged over an approximately Gaussian knowledge state   $\mathcal{P}'(\varphi'|m',\,D')= \mathcal{G}(\varphi-m',D')$ with mean $ m' $ and variance $ D' $. The inference temperature  $T=1$ was kept in the formula to highlight the formal similarity to thermodynamics. Finally,
\begin{eqnarray}
S(D') &=& - \int D\varphi\, \mathcal{G}(\varphi-m',D') \,\ln \left(\mathcal{G}(\varphi-m',D')\right)\nonumber\\
& =&  \frac{1}{2}\, \ln\left| 2\,\pi\,e\,D'\right|
\end{eqnarray}
is the Boltzmann-Shannon entropy of this Gaussian state. It measures the phase space volume of the uncertainty of the approximative knowledge state.

The Gibbs free energy can be rewritten as
\begin{eqnarray}
	G(m',D') &=& \int D\varphi\,  \mathcal{G}(\varphi-m',D')\,
	\ln \left( \frac{\mathcal{G}(\varphi-m',D')}{\exp(-\mathcal{H}(d,\varphi))} \right) \nonumber\\
	&=&  \int D\varphi\,  \mathcal{P}'(\varphi'|m',\,D')\,
	\ln \left( \frac{\mathcal{P}'(\varphi'|m',\,D')}{\mathcal{P}(d,\varphi))} \right) \nonumber\\
    &=&  \int D\varphi\,  \mathcal{P}'(\varphi'|m',\,D')\,
	\ln \left( \frac{ \mathcal{P}'(\varphi'|m',\,D')}{\mathcal{P}(\varphi|d)\, \mathcal{P}(d) } \right) \nonumber\\
	&=&  \mathrm{KL}\left[\mathcal{P}'||\mathcal{P}\right] -\ln \mathcal{Z}(d),
\end{eqnarray}
where $ \mathrm{KL}\left[\mathcal{P}'||\mathcal{P}\right]$ denotes the Kullback-Leibler (KL) divergence of $ \mathcal{P}'(\varphi'|m',\,D') $ and $ \mathcal{P}(\varphi|d) $. Thus, the Gibbs free energy is (up to an irrelevant constant) the information distance between the correct and approximate probability distributions.%
\footnote{The Gibbs free energy is equivalent to $ \mathrm{KL}(\mathcal{P}'||\mathcal{P}) $, the amount of information needed (as measured in nits) to change the correct posterior $ \mathcal{P} $ to the approximate posterior $ \mathcal{P}' $. Information theoretically, the reverse would be more appropriate, as $ \mathrm{KL}(\mathcal{P}||\mathcal{P}') $ would measure how much information has to be added to the approximate posterior $ \mathcal{P}' $ to restore the actual $ \mathcal{P} $ \cite{2017Entrp..19..402L}. This, however, would involve the calculation of non-Gaussian path integrals and is therefore often not an option.}

In case of our toy example, the Gibbs free energy reads
\begin{eqnarray}\label{eq:Gtoy}
G(m',D') &\widehat{=} & \frac{1}{2}\,\mathrm{tr}\left[ \Phi^{-1} \left( m'\,m'^\dagger + D' \right) \right]   + 1^\dagger R\,
e^{3\,m'+ \frac{9}{2}\,\widehat{D'}} \nonumber\\
&&
- d^\dagger \ln \left( R\,e^{3\,m'+ \frac{9}{2}\,\widehat{D'}} \right)
- \frac{1}{2}\, \ln\left| 2\,\pi\,e\,D'\right|,
\end{eqnarray}
since $\langle \varphi \, \varphi^\dagger \rangle_{\mathcal{P}'} = m'\,m'^\dagger + D'$ and $\langle e^\varphi \rangle_{\mathcal{P}'} = e^{m'+\widehat{D'}/2}$. Here $\widehat{D'}_x = D'_{xx}$ denotes the diagonal of $D'$.

The mean field $ m' $ is the minimum of the Gibbs free energy. A short calculation analogously to Eq.~\ref*{eq:MAP} yields
\begin{equation}
 m' =    3\,\Phi\, e^{3\,m''} \,R^\dagger \left(  \frac{d- R\,e^{3\,m''}}{R\,e^{3\,m''}}\right),\label{eq:m'}
\end{equation}
with $m'' = m'+ \frac{3}{2}\,\widehat{D'}$.
The mean field $m'$ therefore depends on the uncertainty $D'$, which itself has to be determined, either by minimizing Eq.~\ref{eq:Gtoy}, or by using equivalently
the thermodynamical relation  \cite{2010PhRvE..82e1112E}
\begin{equation}
D' = \left(\frac{\partial^2 G(m',D') }{\partial m' \partial m'^\dagger}\right)^{-1}.
\end{equation}
The latter yields
\begin{eqnarray}
({D'}^{-1})_{xy} &=&
\Phi^{-1}_{xy} + 9\,\delta_{xy} e^{3\,m''_x} \,R_x^\dagger
\left( 1- \frac{d}{ R\,e^{3\,m''}}\right) \nonumber\\
&&
+ 9 \, e^{3\,m''_x+3\,m''_y} \sum_i \,
 \frac{R_{ix} \, R_{iy}\,d_i}{ (R\,e^{3\,m''})_i^2 },
\end{eqnarray}
a rather complex expression, which depends on $m'$ and $D'$ itself. Thus the mean field and its uncertainty co-variance have to be solved for jointly, as they are interdependent.

A comparison of the mean field and Laplace approximations shows that the estimated uncertainty of the mean field is larger and this estimate is therefore more conservative. The Laplace approximation usually underestimates uncertainties.This larger uncertainty also provides corrections to the (mean) field (with respect to the Laplace/MAP field).

\subsection{Operator formalism}\label{sec:fieldoperator}

The usage of the Gibbs free energy or KL divergence requires the calculation of Gaussian averages of the Hamiltonian. Such Gaussian averages can elegantly be performed using the operator representation IFT. Similar to quantum field theory, a field functional $F(\varphi)$ is averaged over $\mathcal{P}(\varphi|d) = \mathcal{G}(\varphi-m,D)$ with $d=(m,D)$ by inserting the field operator \cite{2016PhRvE..94e3306L},
\begin{equation}
O_m = m + D\, \partial_m
\end{equation}
into the functional,
\begin{equation}
\langle F(\varphi)\rangle_{(\varphi|d)}= F(O_m).
\end{equation}
This operator is set up to generate a field instance out of a Gaussian posterior knowledge state,
\begin{eqnarray}
O_m\,\mathcal{G}(\varphi-m, D) &=&
\frac{m + D\, \partial_m}{\sqrt{|2\pi\,D|}}\,e^{-\frac{1}{2} (\varphi -m)^\dagger D^{-1} (\varphi -m)}
\nonumber\\
&=&
\left( m + D\,  D^{-1} (\varphi -m)  \right)\,\mathcal{G}(\varphi-m, D)
\nonumber\\
&=&
\varphi\,\mathcal{G}(\varphi-m, D).
\end{eqnarray}
This permits to calculate Gaussian expectation values algebraically, since
\begin{eqnarray}\label{eq:operator}
 \langle F(\varphi) \rangle_{(\varphi|d)}
 &=&
 \int \mathcal{D}\varphi \, F[O_m]\,\mathcal{G}(\varphi-m, D) \nonumber\\
 &=& F[O_m]\, \langle 1 \rangle_{(\varphi|d)} =
 F[O_m]\,1 \equiv F[O_m].
\end{eqnarray}
For the usage of this trick, $O_m = m + D\, \partial_m = c + a$ has to be split into a (mean) field creator $c=m$ and a field annihilator $a= D\, \partial_m $ in the expression  $ F[O_m]$. 
Since the field annihilator become zero if they act on the constant functional $1:m\rightarrow 1$, as $c\,1= D\,\partial_m 1 = 0$, they disappear if moved to the very right in the expression of  $ F[O_m]\,1$. This can be achieved with the help of the commutator relations
\begin{equation}
[a_x,c_y]=D_{xy} \mbox{ and }[a_x,a_y]=[c_x,c_y]=0.
\end{equation}
The non-trivial commutation of annihilator and creator generate uncertainty corrections to averages over non-linear moments of the field. For example,
\begin{eqnarray}
\langle \varphi_x \varphi_y\rangle_{(\varphi|d)}&=&  O_{m_x} O_{m_y}\,1=(c_x+a_x)\,(c_y + a_y) \,1\nonumber\\
&=&
c_x c_y 1 + c_x a_y 1 + a_x c_y 1 +a_x a_y 1 \nonumber\\
&=& m_x m_y + 0 + (c_y a_x + [a_x,c_y])\,1 + 0 \nonumber\\
&=& m_x m_y + D_{xy}.
\end{eqnarray}

\subsection{Kullbach-Leibler sampling}

For numerical works, the expressions for the analytically calculated Gibbs free energy or KL divergence might become too complex. As Eq.~\ref{eq:m'}, they often contain diagonals and traces of operators, like $\widehat{D}$ and $\mathrm{Tr}(D)$. These can often only be calculated via sampling techniques, since the operator $D$ is only implicitly defined and only available as a computer routine. The sampling recipes are:
\begin{eqnarray}
\widehat{D} &\approx& \langle \xi\, D\,\xi\rangle_{\{\xi\}} \mbox{ (no }\dagger\mbox{)},\\
\mathrm{Tr}(D)  &\approx& \langle \xi^\dagger\, D\,\xi\rangle_{\{\xi\}}, \mbox{ with}\\
\xi &\hookleftarrow& \mathcal{G}(\xi,\mathbb{1}).
\end{eqnarray}  
Instead of sampling such terms to calculate the expectation value of a Hamiltonian, one can average the Hamiltonian directly over appropriate samples:
\begin{eqnarray}
\langle \mathcal{H}(d,\varphi)\rangle_{\mathcal{G}(\varphi-m,D)} &\approx& \langle  \mathcal{H}(d,\varphi)
\rangle_{\{\varphi\}}, \mbox{ with}\\
\varphi &\hookleftarrow& \mathcal{G}(\varphi -m,D).
\end{eqnarray}  
The drawing of samples from $\mathcal{G}(\varphi - m,D)$ can be done via a virtual signal and data generation in case $D$ has the structure of an information propagator, $D=(\Phi^{-1} + R^\dagger N^{-1} R)^{-1}$:
\begin{eqnarray}
\varphi'  &\hookleftarrow& \mathcal{G}(\varphi',\Phi)\\
n' &\hookleftarrow& \mathcal{G}(n',N)\\
d' &=& R\,\varphi' + n'\\
m' &=& D\,R^\dagger N^{-1} d'\\
\varepsilon' &=& \varphi'-m'\\
\varphi &=& m + \varepsilon'.
\end{eqnarray}
The virtual reconstruction error $\varepsilon' = \varphi'-m'$ obeys the right statistics, as
\begin{eqnarray}
\varepsilon' &=& \varphi'- D\,R^\dagger N^{-1} ( R\,\varphi' + n' )\nonumber\\
&=& D\,\left( D^{-1}- R^\dagger N^{-1}  R\right)\,\varphi' + D\,R^\dagger N^{-1}n' \nonumber\\
&=& D\,\left( \Phi ^{-1} \,\varphi' + R^\dagger N^{-1}n'\right)
\end{eqnarray}
and therefore
\begin{eqnarray}
\langle \varepsilon' \varepsilon'^\dagger \rangle_{(n',\varphi')} &=&  D\, \Phi^{-1} \langle \varphi' \varphi'^\dagger \rangle_{(\varphi')} \Phi^{-1} D\nonumber\\
&&
+ 
D\, R^\dagger N^{-1}
\langle n' n'^\dagger \rangle_{(n')}N^{-1} R\, D\nonumber\\
&=&  D\, \Phi^{-1} D
+ 
D\, R^\dagger N^{-1} R\, D\nonumber\\
&=&  D\, D^{-1} D = D.
\end{eqnarray}
It is important to note that also derivatives, e.g. with respect the mean field, can be expressed via samples,
\begin{eqnarray}
\partial_m \langle \mathcal{H}(d,\varphi)\rangle_{\mathcal{G}(\varphi-m,D)} &\approx&
\int\mathcal{D}\varphi \, \mathcal{H}(d,\varphi) \partial_m\mathcal{G}(\varphi-m,D)\nonumber\\
&=&
- \int \mathcal{D}\varphi \, \mathcal{H}(d,\varphi) \partial_\varphi \mathcal{G}(\varphi-m,D))\nonumber\\
&=&
\int \mathcal{D}\varphi \,\mathcal{G}(\varphi-m,D)   \partial_\varphi \mathcal{H}(d,\varphi) \partial_\varphi\nonumber\\
&=& \langle \partial_\varphi \mathcal{H}(d,\varphi)\rangle_{\mathcal{G}(\varphi-m,D)}\nonumber\\
&=& \langle \partial_\varphi \mathcal{H}(d,\varphi)\rangle_{\{\varphi\}}.
\end{eqnarray}
Such KL-sampling techniques enable the inference of interdependent fields. For example the blind separation of several mixed signals, for which a simple MAP estimate performs very poorly, could be achieved this way  \cite{2017PhRvE..96d2114K}.

\section{Applications}\label{sec:applications}

Current applications of IFT are mostly in astrophysics and cosmology. The elements of the theory presented here already cover many typical measurement situations in these areas: Gaussian and Poissonian noise, Gaussian and log-Gaussian fields, linear instrument responses, known and unknown field correlation structures, etc. A comparison of IFT based methods to existing algorithms can be found in many of the publication  referred below. 

\subsection{Numerical information field theory}

The implementation of algorithms derived within of IFT is facilitated by the \textsc{NIFTy} library \footnote{\url{https://gitlab.mpcdf.mpg.de/ift/NIFTy/}} for \textit{numerical information field theory} \cite{2013A&A...554A..26S, 2017arXiv170801073S}. For example, the calculations for Figs.~\ref{fig:Wiener_filter} and \ref{fig:Poisson} were performed by \textsc{NIFTy} and the corresponding code\footnote{\url{https://gitlab.mpcdf.mpg.de/ift/NIFTy/blob/NIFTy_4/demos/poisson_demo.py}} became part of the current demo package delivered with the library.

\textsc{NIFTy} permits the abstract implementation of IFT formulas irrespective of the underlying domains the fields live over. The same algorithm implemented with  \textsc{NIFTy} can reconstruct without change fields living over 1D, 2D, 3D Cartesian spaces, the sphere, or even product spaces built out of those spaces. \textsc{NIFTy} takes care of many details of the space pixelization, the operator calculus, the set up of correlation structures,  harmonic transforms of fields, and numerical minimization schemes. This permits the user to concentrate on the essentials of her information field theoretical problem and less on implementation details. Most numerical operations are performed by standard libraries. 

With \textsc{NIFTy}, different algorithmic choices for signal estimation are possible. Algorithms can be based on MAP or Gibbs-free energy minimization, or use perturbation series, like expansion in Feynman diagrams. For the joint inference of a signal and its power spectrum the mean field approach of \cite{2017arXiv171102955K} is performing well.

In the following we highlight a number of applications of IFT to illustrate the spectrum of potential usages: photon count imaging, other available imaging and signal reconstruction methods, non-Gaussianity estimation, and simulation of field dynamics.

\subsection{Photon imaging}

Photon imaging is the reconstruction of the continuous spatial and/or spectral photon emission field from which a finite number of detected photons were emitted. The D$^3$PO and D$^4$PO algorithms \cite{2015A&A...574A..74S, 2018arXiv180202013P} address this task. The first one\footnote{\url{http://ascl.net/1504.018}},  D$^3$PO, denoises, deconvolves and decomposes photon observations (thus the acronym  D$^3$PO) into a diffuse log-emission field  plus a point-source log-emission field. For the diffuse field a power spectrum is inferred as well. For the point source field point sources are assumed to reside in every image pixel, just most of them being insignificantly faint. The application of D$^3$PO to data of the Fermi gamma-ray satellite is shown in Fig.~\ref{fig:fermi-diffuse}  \cite{2015A&A...581A.126S}.

The successor algorithm  D$^4$PO does essentially the same, just with the difference that the fields can be defined on product spaces of spatial coordinates (the sky) and other domains (like the photon energy space). Furthermore, an arbitrary number of such fields can be reconstructed simultaneously, e.g. to allow also for background counts in data space. The correlation structure of the fields is assumed to be the direct product over (a priori unknown and therefore simultaneously reconstructed) correlation functions in the different directions. A diffuse log-emission field $\varphi(x,y)$ over the product space of sky ($x\in S^2$) and log-energy ($y=\ln (E/E_0) \in \mathbb{R}$) therefore has the correlation structure $ \Phi_{(x,y)\,(x',y')}=\Phi^{(S^2)}_{xx'}\,\Phi^{(E)}_{yy'} $.

\subsection{Available IFT algorithms}

\begin{figure*}
	\centering
	\includegraphics[width=\linewidth]{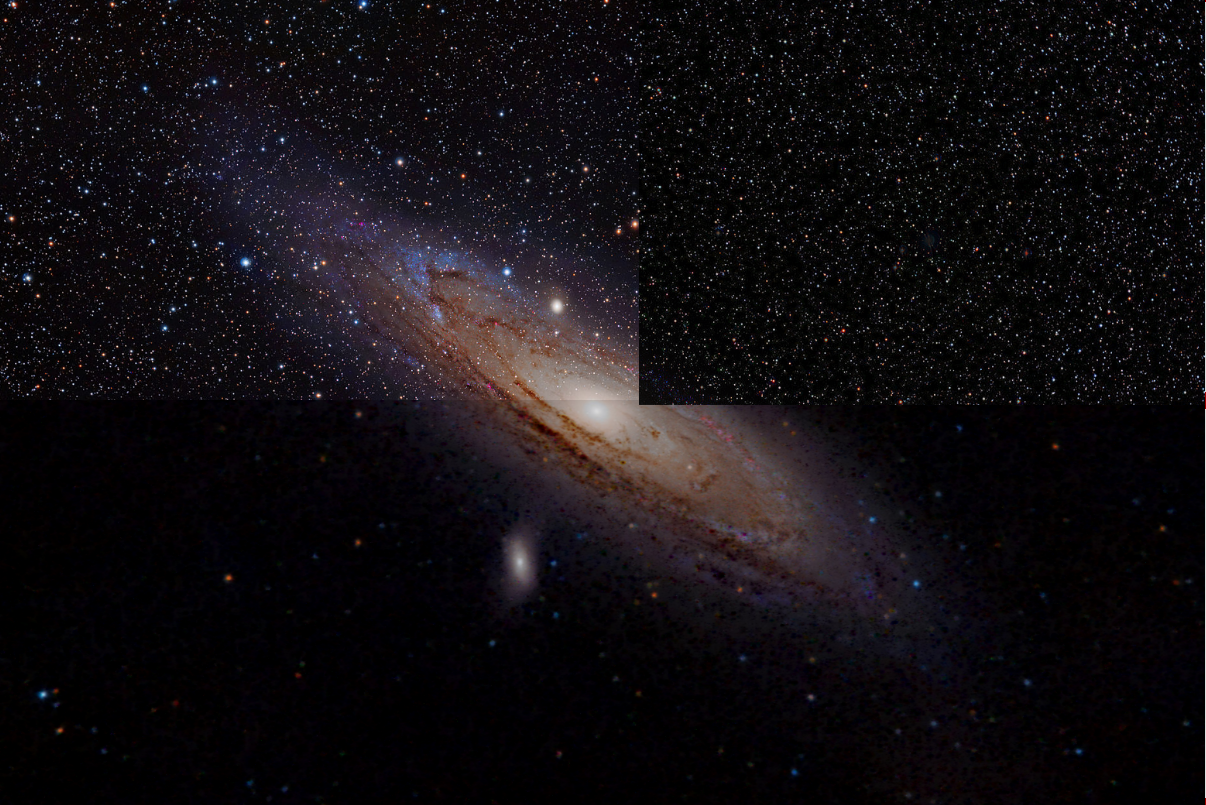}
	\caption{
		Hubble Space Telescope image of the Andromeda galaxy (top left for original image) separated by starblade into stars (top right) and diffuse emission (bottom). Here, an early version of starblade was applied separately to the RGB channels of a JPEG image \cite{wiki:Andromeda} for illustration purpose only. Figure kindly provided by Jakob Knollm\"uller, the author of starblade.
	}
	\label{fig:starblade}
\end{figure*}

A number of further IFT algorithms are already freely available as open source codes. Here an incomplete list:
\begin{itemize}
	\item[\textbf{RESOLVE}]
	images the radio sky from interferometric data  \cite{2015A&A...581A..59J, 2016A&A...586A..76J, 2016arXiv160504317G, 2018arXiv180302174A}.
	Interferometers measure individual Fourier components of the brightness field and a reconstruction is indispensable to obtain an image. Despite the very different instrument response and noise statistics it deals with, RESOLVE is similar to D$^3$PO and D$^4$PO in that it describes diffuse emission as a log-normal field with unknown power spectrum.
	\item[\textbf{charm}] reconstructs non-parametrically the expansion history of the Universe from supernovae 1a data \cite{2017ascl.soft03015P}. 
	It has confirmed the concordance of this dataset with the $\Lambda$CDM model \cite{2017A&A...599A..92P}.
	\item[\textbf{PySEA}] is an open-source project dedicated to provide a generic Python framework for spatially explicit statistical analyses of point clouds and other geospatial data, in the spatial and frequency domains, for use in the geosciences.\footnote{Quote from \url{https://github.com/dbuscombe-usgs/pysesa}.}
	\item[\textbf{starblade}] separates point sources from diffuse emission while estimating and using the correlation structure of the latter (Fig.~\ref{fig:starblade}) \cite{2018arXiv180405591K}. It assumes the data to be noiseless and is therefore intended for the post-processing of high fidelity images or as an internal processing step in other imaging codes.
	\end{itemize}
	
\subsection{Non-Gaussianity}

An important question of contemporary cosmology is how precisely the initial density fluctuations of the Universe were following Gaussian statistics. The different inflationary scenarios for the first fractions of a second of the cosmic history predict different amounts of deviations from Gaussianity. A simple parametrisation of non-Gaussianity is the so-called $ f_\mathrm{nl} $ model, in which an initially Gaussian random field $ \varphi\leftarrow \mathcal{G}(\varphi,\Phi) $ is non-linearly processed into the gravitational potential in the Early Universe  observed at a later epoch:
\begin{equation}
\psi(x) = \psi[\varphi, f_\mathrm{nl}](x) = \varphi(x) + f_\mathrm{nl} \,
\left[\varphi^2(x)- \langle \varphi^2(x)\rangle_{(\varphi)}\right]
\end{equation}
If the measurement process is linear and Gaussian, with $ d=R\,\psi+n $ and $ n\leftarrow \mathcal{G}(n,N) $, the information Hamiltonian
\begin{eqnarray}
\mathcal{H}(d,\varphi|f_\mathrm{nl}) &\widehat{=}&
\frac{1}{2} \, \left( d - R \,\psi[\varphi, f_\mathrm{nl}] \right)^\dagger \, N^{-1}
\left( d - R \, \psi[\varphi, f_\mathrm{nl}] \right)
\nonumber\\
&& +
\frac{1}{2} \, \varphi^\dagger \Phi \,\varphi  \label{eq:fnl}
\end{eqnarray}
becomes fourth order in $\varphi$.

In order to decide which value  $ f_\mathrm{nl} $ the data prefer, one needs to calculate the evidence $\mathcal{P}( d | f_\mathrm{nl} )$. This is, however, also the partition function $\mathcal{Z} (d | f_\mathrm{nl} )$. Since $f_\mathrm{nl} $ is a (comparably) small parameter, one can use that the logarithm of the partition function is given by the sum of all simply connected Feynman diagrams without external vertices. Sorting such diagrams by their order in  $f_\mathrm{nl} $ up to second order, one can construct the MAP estimator for  $f_\mathrm{nl} $ \cite{2009PhRvD..80j5005E}
. This is displayed in  Fig.~\ref{fig:fnl} superimposed to a map of the CMB, to which such a non-Gaussianity estimator can be applied. The Feynman diagrams containing  $f_\mathrm{nl} $ up to linear order are in the numerator and the ones up to quadratic order in the denominator of the fraction that comprises the  $f_\mathrm{nl} $-estimator (while $f_\mathrm{nl} $ was removed from these terms as it is to be estimated). The numerator is identical to the well known Komatsu-Spergel-Wandelt (KSW) estimator \cite{2005ApJ...634...14K} used in CMB studies, the denominator provides a Bayesian normalization, which depends on the particular data realization.
Actually, the original KSW estimator comprised only the terms given by the first two diagrams of the numerator. The other terms were discovered later as correction terms for inhomogeneous noise \cite{2006JCAP...05..004C}.
The diagrammatic approach to IFT delivers all these terms naturally and simultaneously. Further details on this can be found in \cite{2009PhRvD..80j5005E}.

\subsection{Information field dynamics}

As a final application, we show how IFT may be used in future to build computer simulation schemes for partial differential equations (PDEs). A field $ \varphi(x,t)$ over space and time may follow a PDE of the form
\begin{equation}
\partial_t \varphi = F[\varphi].
\end{equation}
For example $ F[\varphi] =  \kappa\, \partial^2_x \varphi + \nu^{1/2}\, \xi$ reproduces Eq.~\ref{eq:pde}.

On a computer, only a discretized version of the field can be stored in memory. In the framework of information field dynamics (IFD, \cite{2013PhRvE..87a3308E, 2017arXiv170902859L, 2018arXiv180200971D}), the data vector in computer memory describing the field at an instant $t$  is regarded as the result of a virtual measurement process of the form given by Eq.~\ref{eq:d}. A convenient linear response $R$ might be given by the pixel window function, e.g. $R_{ix} = P(x\in \Omega_i|x,\Omega_i)\in \{0,1\}$ such that $R_{ix} =1 $ if $x$ is within the volume of the $i$-th pixel  $\Omega_i$, otherwise $R_{ix} =0 $. In this case the data would contain the pixel-integrated field values. The measurement noise might be absent, $n=0$, as the virtual measurements can be chosen to be noiseless.

However, even without measurement uncertainty, the field is not fully determined by the data. The remaining uncertainty is captured by the field posterior $\mathcal{P}(\varphi|d)$, which can be specified via Bayes' theorem (Eq.~\ref{eq:BayesH}) in case a field prior $\mathcal{P}(\varphi)$ is available.

In essence, IFD evolves the knowledge on the field as parametrized by the posterior by evolving the data $d$ in such a way that information on the field is conserved as much as possible. For this, the KL-divergece between the time evolved field posteriors and a posterior from later time data $ d' $ is minimized.

Thus, the idea of IFD is to find an evolution equation for data $d$ in computer memory, which captures and incorporates the fully resolved field evolution and the prior knowledge as well as possible. As this data vector is finite dimensional, its evolution equation is then only an \textit{ordinary differntial equations} (ODE), which can be solved with standard numerical methods. The data, however, implies a PDF over the full infinite dimensional field configuration space, which can be questioned for any quantity of interest. 

\begin{figure}
	\centering
	\includegraphics[width=\linewidth, trim={0cm 0cm 0cm 0},clip]{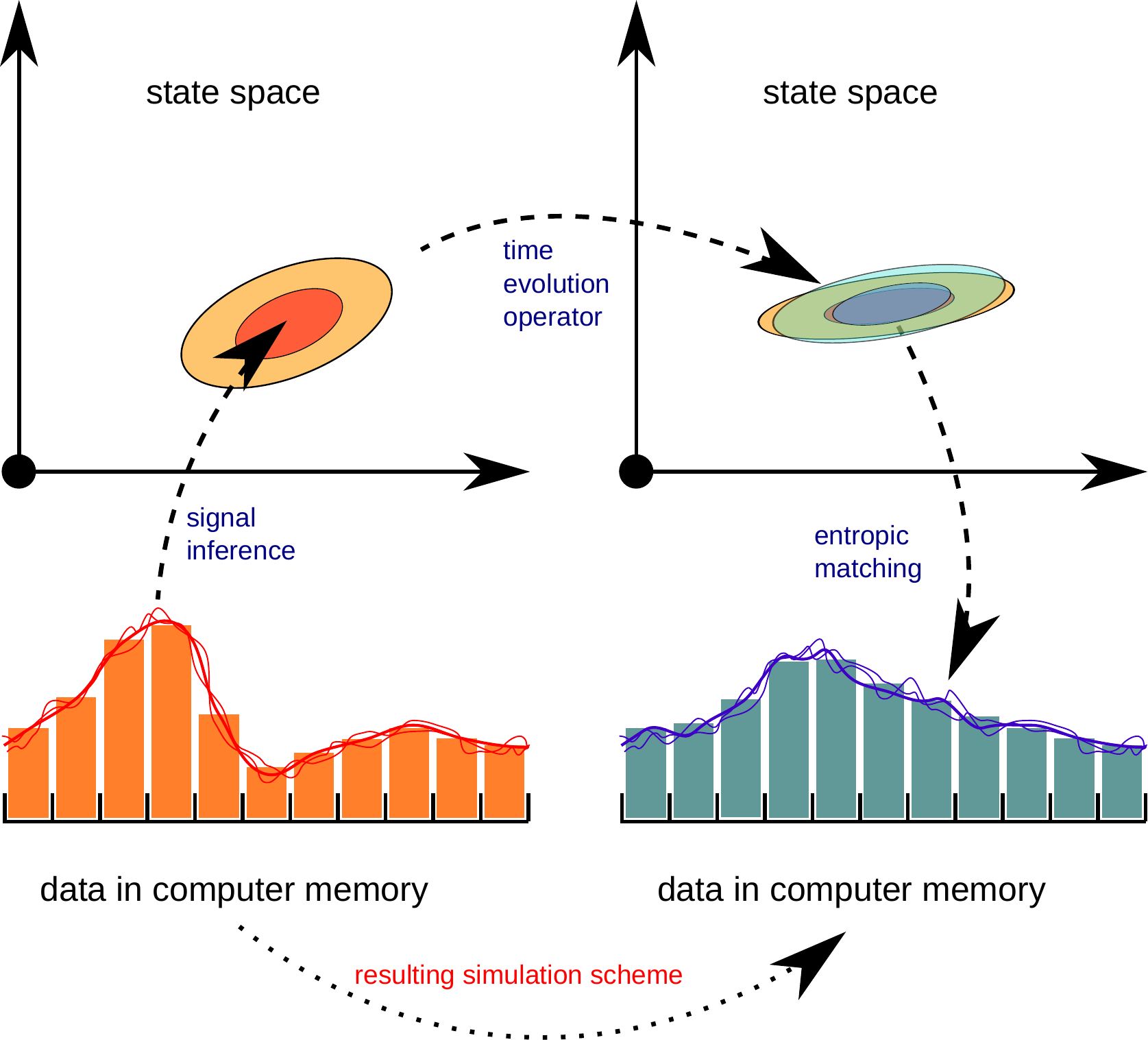}
	\caption{Sketch of the construction of simulation schemes via IFD. The data in computer memory (orange bars) at some time imply a posterior in field configuration space (orange contours). Each point of this PDF (e.g. orange sample curves shown with data) is a continuous field that can be evolved mathematically  to a future time according to the PDE to be simulated. The evolved state has to be represented by data again. This is chosen via entropic matching, such that the new data contains as much information as possible about the evolved field. The combination of these operations defines a data space-only operation, the simulation scheme, which can be performed by a computer. It incorporates the data, prior knowledge on the field, and the dynamical laws in an information optimal way.}
	\label{fig:IFD}
\end{figure}

Let us regard the case in which the field prior is Gaussian, $\mathcal{P}(\varphi)= \mathcal{G}(\varphi, \Phi)$, the measurement is linear and noise-free, $ d=R\,\varphi $, and thus the posterior is Gaussian as well $ \mathcal{P}(\varphi|d)=  \mathcal{G}(\varphi-m, D) $, with $m= m(d) =  \Phi \,R^\dagger (R\,\Phi\,R^\dagger)^{-1} d \equiv W\,d$ is the noise-less Wiener filter reconstruction.
In this case, the optimal temporal evolution of the data is given by \cite{2017arXiv170902859L}
\begin{equation}\label{eq:IFD0}
\partial_t d = R\, \langle F(\varphi) \rangle_{(\varphi|d)}.
\end{equation}
The time derivative of the data is given by the measurement response $R$ applied to the time derivative of the field, averaged over all posterior field configurations. This gives now an ordinary differential equation for the data, which can be solved with appropriate numerical methods. The evolution of the data implies an evolving field probability density $\mathcal{P}(\varphi|d(t))$, which encodes the knowledge about the field at all instances, including its mean value and its uncertainty.

Since the measurement equation was chosen to be linear, and the field prior to be Gaussian, the posterior will be Gaussian as well. Nevertheless, the $F(\varphi)$ term is non-linear for a non-linear PDE and averaging it might be non-trivial analytically. 

The field operator trick from Sect.~\ref{sec:fieldoperator} helps here.  Since the average is over a Gaussian knowledge state, this expression can be brought into the compact form
\begin{equation}\label{eq:IFD}
\partial_t d = R\,F[O_m],
\end{equation}
where
\begin{equation}
O_m = m + D\, \partial_m
\end{equation}
is the field operator.

Eq.~\ref{eq:IFD} describes the data evolution that captures the (knowledge on the) field evolution best.
For a linear PDE, like Eq.~\ref{eq:pde}, the $\partial_m$ terms make no difference, and the data equation becomes
\begin{equation}\label{eq:IFDl}
\partial_t d = R\,F[m] =
\underbrace{R\,F\,\Phi\,R^\dagger (R\,\Phi\,R^\dagger)^{-1}}_{\equiv \widetilde{F}}
  d = \widetilde{F}\,d,
\end{equation}
a linear ODE. The data evolution operator $\widetilde{F}$ contains a weighting of the action of the PDE operator $F$ with the statistics of the expected field fluctuations. These are encoded in $\Phi$ and its projection into the data space via the measurement response $R$. This weight $R\,\Phi\,R^\dagger$ is also the `denominator' that ensures the proper units of the data space operator. This way, the knowledge about the field modes, which are not resolved by the data, enters the data dynamics. For example, an IFD algorithm for a thermally excited Klein-Gordon field performs slightly better than a spectral scheme for the same problem, because the former exploits its knowledge about the correct statistics of the unresolved field fluctuations \cite{2013PhRvE..87a3308E}.

For a non-linear PDE, quadratic terms in $F$ generate quadratic terms in the field operator, $ O^2_m =  (m + D\, \partial_m)\, (m + D\, \partial_m) = m^2 + D + \mathcal{O}(\partial_m)$ that contain non-vanishing contributions
from the uncertainty dispersion $ D $. Thus, the data of a non-linearly evolving field not only get a non-linear ODE, as $m^2=(W\,d)^2$ is quadratic in the data, they also get corrections that capture the effect of the uncertainty processed through the non-linearity, expressed by terms that contain the field uncertainty $D$.

A naive discretization of a PDE, which just replaces the differential operators of the PDE with difference operators on the data, does not account for such non-linear sub-grid effects. For example, IFD schemes for the Burgers equation, which is a simplistic version of compressive hydrodynamics,  handle the shock waves developing in the field dynamics better than (central) finite difference schemes \cite{2017arXiv170902859L}, the latter representing the (most) naive discretization of the PDE.

Despite these encouraging results, it should be noted that IFD requires further theoretical investigation and algorithmic work before it can seriously compete with existing state-of-the art simulation schemes in terms of usability and performance. For example, how the field prior knowledge should evolve according to the PDE to be simulated is currently an open question.

Anyhow, IFD opens the door for advanced simulations. For example, as IFD is already formulated probabilistically, the assimilation of observational data into an IFD simulation may be relatively straight-forward.

\section{Outlook}\label{sec:outlook}
IFT, the information theory for fields, has many potential scientific, technological, and economical applications. Two current development lines to bring IFT into broader usage are presented as a closing outlook:

\textbf{Imaging}, the transformation of astronomical or medical data into visual informative images, is a central application area of IFT. The recipe to an IFT imaging algorithm is the construction of the information Hamiltonian and/or its Gibbs free energy, and the minimization of those with respect to the unknown fields. The joint Hamiltonian is comprised of an instrument description $ \mathcal{H}(d|\varphi,\theta) $, a field prior $ \mathcal{H}(\varphi|\theta) $, and hyper-prior $ \mathcal{H}(\theta) $ of all the remaining unknowns $ \theta  $ (field or noise power spectra, calibration parameters of the response, etc.). Thus, imaging of very different instruments can be brought into a unified description, in which even the data of different instruments can be imaged jointly. To this end, an \textit{Universal Bayesian Imaging toolKit} (UBIK)%
\footnote{The name derives from the novel UBIK by Phillip K. Dick (1969), in which an agent called UBIK is essential to restore reality whenever this gets corrupted.}
is under development, which will permit imaging based on multi-instrument data of fields living over multi-dimensional spaces with spatial, spectral, and/or temporal dimensions.

Besides the already described IFD development, the \textbf{inference of dynamical fields} including their unknown dynamics from data is a relevant research direction with promising first results \cite{2017PhRvE..96e2104F}. Observations of an evolving field might be sufficient to determine the dynamical laws the field obeys, which then can be used to better estimate and predict the field from limited observational data.

IFT not always provides novel methods. In many cases it just reproduces well-established algorithms. The Wiener filter is such an example, which emerges in IFT as the simplest possible inference algorithm in case of linear measurement and independent Gaussian signal and noise statistics. Thus, IFT can shed light on the conditions under which an existing method becomes optimal. Furthermore, IFT permits to extend such methods consistently to more complex, non-linear, and non-Gaussian situations. These may perform superior to traditional approaches, in case the assumptions used in the IFT algorithm derivation are sufficiently met by the actual measurement situation at hand.

To summarize, IFT seems to be applicable within all areas in which fields have to be handled under uncertainty. IFT is the appropriate language for field inference from imperfect data. Currently it is mostly used in astrophysics and cosmology, however, its potential for geophysics, medical imaging, and other areas of remote and non-invasive sensing should be obvious. 

This short introduction to IFT aimed to show that  physicists have the very powerful language of field theory at their disposal to address field inference problems. Whether the  resulting method performs better or worse than traditional approaches depends on whether correct assumptions were made or not. In any case, an IFT-based investigation of an inference problem will provide deeper insights into the nature of the available information on the field of interest.

\section*{Acknowledgments}

The work described here would not have been possible without the support, inputs, ideas, criticism, discussion, and work of many collaborators, students, and friends. In particular among them I like to thank 
Michael Bell, Martin Dupont, Philipp Frank, Mona Frommert, Maxim Greiner, Sebastian Hutschenreuter, Henrik Junklewitz, Francisco-Shu Kitaura, Jakob Knollmüller, Reimar Leike, Christoph Lienhard, Niels Oppermann, Natalia Porqueres, Daniel Pumpe, Martin Reinecke, Georg Robbers, Marco Selig, Theo Steininger, Valentina Vacca, and Cornelius Weig for their valuable contributions to the field, which I had the honor to proudly present here. I want to appologize to all the others, not mentioned here, for me not including their work into this article. The presentation of this work benefited from detailed feedback by Jakob Knollm\"uller, Reimar Leike, Ancla Müller, Martin Reinecke, Kandaswamy Subramanian, and two anonymous referees.

\bibliographystyle{andp2012}
\bibliography{tae}

\providecommand{\WileyBibTextsc}{}
\let\textsc\WileyBibTextsc
\providecommand{\othercit}{}
\providecommand{\jr}[1]{#1}
\providecommand{\etal}{~et~al.}


\begin{thebibliography}{[10]}

\bibitem{2009PhRvD..80j5005E}
 \textsc{T.\,A. {En{\ss}lin}},  \textsc{M.~{Frommert}},  and  \textsc{F.\,S.
  {Kitaura}} \jr{\prd} \textbf{80}(10), 105005 (2009).


\othercit
\bibitem{2013AIPC.1553..184E}
 \textsc{T.~{En{\ss}lin}},
{Information field theory},
 in: American Institute of Physics Conference Series, edited by U.~{von
  Toussaint}, , American Institute of Physics Conference Series Vol.\,1553
  (August 2013),  pp.\,184--191.


\bibitem{2014AIPC.1636...49E}
 \textsc{T.~{En{\ss}lin}} \jr{Bayesian Inference and Maximum Entropy Methods in
  Science and Engineering} \textbf{1636}(December), 49--54 (2014).


\othercit
\bibitem{wiki:xxx}
 \textsc{W.~contributors},
Information field theory --- wikipedia{,} the free encyclopedia, 2018,
[Online; accessed 3-February-2018].


\othercit
\bibitem{Lemm2003}
 \textsc{J.\,C. {Lemm}},
Bayesian Field Theory (Johns Hopkins University Press, 2003).


\othercit
\bibitem{1982ieee...70..939J}
 \textsc{E.\,T. {Jaynes}},
{On the Rationale of Maximum Entropy Methods},
 in: Proc. IEEE, Volume 70, p. 939-952,  (1982),  pp.\,939--952.


\bibitem{2018arXiv180702709L}



\othercit
\bibitem{1949wiener-short}
 \textsc{N.~{Wiener}},
{Extrapolation, Interpolation, and Smoothing of Stationary Time Series} (NY:
  Wiley, 1949).


\bibitem{2011PhRvD..83j5014E}
 \textsc{T.\,A. {En{\ss}lin}} and  \textsc{M.~{Frommert}} \jr{\prd}
  \textbf{83}(10), 105014 (2011).


\bibitem{2013PhRvE..87c2136O}
 \textsc{N.~{Oppermann}},  \textsc{M.~{Selig}},  \textsc{M.\,R. {Bell}},  and
  \textsc{T.\,A. {En{\ss}lin}} \jr{\pre} \textbf{87}(3), 032136 (2013).


\bibitem{2011PhRvE..84d1118O}
 \textsc{N.~{Oppermann}},  \textsc{G.~{Robbers}},  and  \textsc{T.\,A.
  {En{\ss}lin}} \jr{\pre} \textbf{84}(4), 041118 (2011).


\bibitem{2017arXiv171102955K}
 \textsc{J.~{Knollm{\"u}ller}},  \textsc{T.~{Steininger}},  and  \textsc{T.\,A.
  {En{\ss}lin}} \jr{ArXiv e-prints}(November) (2017).


\bibitem{1987PhLB..195..216D}
 \textsc{S.~{Duane}},  \textsc{A.\,D. {Kennedy}},  \textsc{B.\,J. {Pendleton}},
   and  \textsc{D.~{Roweth}} \jr{Physics Letters B} \textbf{195}(September),
  216--222 (1987).


\bibitem{2017arXiv170102434B}
 \textsc{M.~{Betancourt}} \jr{ArXiv e-prints}(January) (2017).


\bibitem{2017Entrp..19..402L}
 \textsc{R.~{Leike}} and  \textsc{T.~{En{\ss}lin}} \jr{Entropy}
  \textbf{19}(August), 402 (2017).


\bibitem{2010PhRvE..82e1112E}
 \textsc{T.\,A. {En{\ss}lin}} and  \textsc{C.~{Weig}} \jr{\pre} \textbf{82}(5),
  051112 (2010).


\bibitem{2016PhRvE..94e3306L}
 \textsc{R.\,H. {Leike}} and  \textsc{T.\,A. {En{\ss}lin}} \jr{\pre}
  \textbf{94}(5), 053306 (2016).


\bibitem{2017PhRvE..96d2114K}
 \textsc{J.~{Knollm{\"u}ller}} and  \textsc{T.\,A. {En{\ss}lin}} \jr{\pre}
  \textbf{96}(4), 042114 (2017).


\bibitem{2013A&A...554A..26S}
 \textsc{M.~{Selig}},  \textsc{M.\,R. {Bell}},  \textsc{H.~{Junklewitz}},
  \textsc{N.~{Oppermann}},  \textsc{M.~{Reinecke}},  \textsc{M.~{Greiner}},
  \textsc{C.~{Pachajoa}},  and  \textsc{T.\,A. {En{\ss}lin}} \jr{\aap}
  \textbf{554}(June), A26 (2013).


\bibitem{2017arXiv170801073S}
 \textsc{T.~{Steininger}},  \textsc{J.~{Dixit}},  \textsc{P.~{Frank}},
  \textsc{M.~{Greiner}},  \textsc{S.~{Hutschenreuter}},
  \textsc{J.~{Knollm{\"u}ller}},  \textsc{R.~{Leike}},
  \textsc{N.~{Porqueres}},  \textsc{D.~{Pumpe}},  \textsc{M.~{Reinecke}},
  \textsc{M.~{{\v S}raml}},  \textsc{C.~{Varady}},  and
  \textsc{T.~{En{\ss}lin}} \jr{ArXiv e-prints}(August) (2017).


\bibitem{2015A&A...574A..74S}
 \textsc{M.~{Selig}} and  \textsc{T.\,A. {En{\ss}lin}} \jr{\aap}
  \textbf{574}(February), A74 (2015).


\bibitem{2018arXiv180202013P}
 \textsc{D.~{Pumpe}},  \textsc{M.~{Reinecke}},  and  \textsc{T.\,A.
  {En{\ss}lin}} \jr{ArXiv e-prints}(February) (2018).


\bibitem{2015A&A...581A.126S}
 \textsc{M.~{Selig}},  \textsc{V.~{Vacca}},  \textsc{N.~{Oppermann}},  and
  \textsc{T.\,A. {En{\ss}lin}} \jr{\aap} \textbf{581}(September), A126 (2015).


\bibitem{2015A&A...581A..59J}
 \textsc{H.~{Junklewitz}},  \textsc{M.\,R. {Bell}},  and
  \textsc{T.~{En{\ss}lin}} \jr{\aap} \textbf{581}(September), A59 (2015).


\bibitem{2016A&A...586A..76J}
 \textsc{H.~{Junklewitz}},  \textsc{M.\,R. {Bell}},  \textsc{M.~{Selig}},  and
  \textsc{T.\,A. {En{\ss}lin}} \jr{\aap} \textbf{586}(February), A76 (2016).


\bibitem{2016arXiv160504317G}
 \textsc{M.~{Greiner}},  \textsc{V.~{Vacca}},  \textsc{H.~{Junklewitz}},  and
  \textsc{T.\,A. {En{\ss}lin}} \jr{ArXiv e-prints}(May) (2016).


\bibitem{2018arXiv180302174A}
 \textsc{P.~{Arras}},  \textsc{J.~{Knollm{\"u}ller}},
  \textsc{H.~{Junklewitz}},  and  \textsc{T.\,A. {En{\ss}lin}} \jr{ArXiv
  e-prints}(March) (2018).


\othercit
\bibitem{2017ascl.soft03015P}
 \textsc{N.~{Porqueres}} and  \textsc{T.\,A. {Ensslin}},
{Charm: Cosmic history agnostic reconstruction method},
Astrophysics Source Code Library, March 2017.


\bibitem{2017A&A...599A..92P}
 \textsc{N.~{Porqueres}},  \textsc{T.\,A. {En{\ss}lin}},
  \textsc{M.~{Greiner}},  \textsc{V.~{B{\"o}hm}},  \textsc{S.~{Dorn}},
  \textsc{P.~{Ruiz-Lapuente}},  and  \textsc{A.~{Manrique}} \jr{\aap}
  \textbf{599}(March), A92 (2017).


\bibitem{2018arXiv180405591K}
 \textsc{J.~{Knollm{\"u}ller}},  \textsc{P.~{Frank}},  and  \textsc{T.\,A.
  {En{\ss}lin}} \jr{ArXiv e-prints}(April) (2018).


\othercit
\bibitem{wiki:Andromeda}
 \textsc{{Wikipedia contributors}},
Andromeda galaxy --- {Wikipedia}{,} the free encyclopedia,
\url{https://en.wikipedia.org/w/index.php?title=Andromeda_Galaxy&oldid=855593236},
  2018,
[Online; accessed 28-August-2018].


\bibitem{2005ApJ...634...14K}
 \textsc{E.~{Komatsu}},  \textsc{D.\,N. {Spergel}},  and  \textsc{B.\,D.
  {Wandelt}} \jr{\apj} \textbf{634}(November), 14--19 (2005).


\bibitem{2006JCAP...05..004C}
 \textsc{P.~{Creminelli}},  \textsc{A.~{Nicolis}},  \textsc{L.~{Senatore}},
  \textsc{M.~{Tegmark}},  and  \textsc{M.~{Zaldarriaga}} \jr{\jcap}
  \textbf{5}(May), 004 (2006).


\bibitem{2013PhRvE..87a3308E}
 \textsc{T.\,A. {En{\ss}lin}} \jr{\pre} \textbf{87}(1), 013308 (2013).


\bibitem{2017arXiv170902859L}
 \textsc{R.\,H. {Leike}} and  \textsc{T.\,A. {En{\ss}lin}} \jr{ArXiv
  e-prints}(September) (2017).


\bibitem{2018arXiv180200971D}
 \textsc{M.~{Dupont}} and  \textsc{T.~{En{\ss}lin}} \jr{ArXiv
  e-prints}(February) (2018).


\bibitem{2017PhRvE..96e2104F}
 \textsc{P.~{Frank}},  \textsc{T.~{Steininger}},  and  \textsc{T.\,A.
  {En{\ss}lin}} \jr{\pre} \textbf{96}(5), 052104 (2017).


\end{thebibliography}
\end{document}